\documentclass[a4paper,12pt]{article}

\usepackage{amsmath}
\usepackage{wasysym}
\usepackage{bm} % Include bold math: \bm{} creates bold letters symbols in math mode
\newcommand{\del}{{\delta}}

\newcommand{\kap}{{\kappa}}

\newcommand{\eqand}{{\quad \mathrm{and} \quad}}

\newcommand{\eth}{{\textrm{\dh}}}
\newcommand{\tho}{{\textrm{\thorn}}}

\newcommand{\lb}{{\ell}}
\newcommand{\nb}{{n}}
\newcommand{\mb}[1]{{m_{(#1)}}}
\newcommand{\eb}{{e}}
\newcommand{\M}[1]{{\stackrel{#1}{M}}}  % In iopart class, this should read {{\overset{#1}{M}}}

\def \bea { \begin{eqnarray}}
\def \eea {\end{eqnarray}}
\def \be {\begin{equation}}
\def \ee {\end{equation}}

\def \bmd #1 {\mb{#1}}

\def \hbm #1 {{\hat \mb{#1}}}

\def\scalar{{\mathbf S}} 
\def\vector{{\mathbf V}} 
\def\tensor{{\mathbf T}} 

\numberwithin{equation}{section}

\title{Algebraically special perturbations of the Schwarzschild solution in higher dimensions}

\author{\'Oscar J. C. Dias${}^{\dagger a}$,   Harvey S. Reall${}^{\ast }$\\ \\
{\small  ${}^{\dagger}$ Institut de Physique Th\'eorique, CEA Saclay,}\\
 {\small CNRS URA 2306, F-91191 Gif-sur-Yvette, France}\\ 
{\small ${}^{\ast}$ DAMTP, Centre for Mathematical Sciences, University of Cambridge}, \\{\small 
 Wilberforce Road, Cambridge CB3 0WA, UK} \\ \\
 \small{ oscar.dias@cea.fr,  H.S.Reall@damtp.cam.ac.uk }}

\begin{document} 
\maketitle

\begin{abstract}
%%%%%%%%%%%
We study algebraically special perturbations of a generalized Schwarzschild solution in any number of dimensions. There are two motivations. First, to learn whether there exist interesting higher-dimensional algebraically special solutions beyond the known ones. Second, algebraically special perturbations present an obstruction to the unique reconstruction of general metric perturbations from gauge-invariant variables analogous to the Teukolsky scalars and it is desirable to know the extent of this non-uniqueness. In four dimensions, our results generalize those of Couch and Newman, who found infinite families of time-dependent algebraically special perturbations. In higher dimensions, we find that the only regular algebraically special perturbations are those corresponding to deformations within the Myers-Perry family. Our results are relevant for several inequivalent definitions of ``algebraically special".
\end{abstract}

%\newpage
%\tableofcontents

%%%%%%%%%%%%%%%%%%%%%%%%%%%%%%%%%%%%%%%%%%%%%
\section{Introduction}\label{sec:intro}
%%%%%%%%%%%%%%%%%%%%%%%%%%%%%%%%%%%%%%%%%%%%%%%%%%

The study of spacetimes with algebraically special Weyl tensor played an important role in the discovery of some important solutions of the 4-dimensional Einstein equation, e.g., the Kerr solution \cite{Kerr:1963ud} and the spinning C-metric \cite{Kinnersley:1969zza}. It seems worthwhile investigating whether the algebraically special property is useful for finding new explicit solutions of the Einstein equation in higher dimensions. In this paper we will consider the vacuum Einstein equation, allowing for a cosmological constant.

Several inequivalent definitions of ``algebraically special" have been proposed in higher dimensions \cite{desmet,cmpp,taghavi,Durkee:2010xq}. (See Ref. \cite{reall} for an introductory review or Ref. \cite{Ortaggio:2012jd} for a more detailed review.) The Myers-Perry black hole solution  \cite{myersperry} is algebraically special according to any of these definitions. However, it is not known whether the class of algebraically special spacetimes in higher dimensions is as rich as in four dimensions. To investigate this, we will study algebraically special {\it perturbations} of a known solution, namely the higher-dimensional Schwarzschild solution. From this we can learn about algebraically special solutions which admit the Schwarzschild solution as a limit. 

A second motivation for this project comes from work on linearized gravitational perturbations of algebraically special solutions such as the Myers-Perry black hole. Ref. \cite{Durkee:2010qu} showed that there exists a gauge-invariant quantity $\delta \Omega_{ij}$ (defined below) which depends linearly and locally on the metric perturbation and has the same number of physical degrees of freedom as the metric perturbation. $\delta \Omega_{ij}$ is the higher-dimensional analogue of one of the Teukolsky scalars used in the study of gravitational perturbations of a Kerr black hole \cite{Teukolsky:1972my}. Knowing $\delta \Omega_{ij}$ determines the metric perturbation up to addition of a solution of $\delta \Omega_{ij}=0$. As we will see, this equation is closely related to the condition for the perturbation to be algebraically special. Hence by determining such perturbations we can identify the extent to which $\delta \Omega_{ij}$ fails to characterize metric peturbations. (For perturbations of a Kerr black hole, this problem was studied in Ref. \cite{wald}.) 

In 4d, algebraically special perturbations of the Schwarzschild solution were studied by Couch and Newman \cite{Couch:1973zc}. They decomposed perturbations into harmonics on $S^2$, labelled by $l=0,1,\ldots$. For $l=0,1$ there are algebraically special perturbations corresponding to a change in the mass, and to the Kerr solution linearized for small angular momentum. For each $l>1$ there are time-dependent algebraically special perturbations of two types. The first type decays exponentially as a function of the retarded time coordinate $u$, and vanishes on the future horizon. This corresponds to the linearization about the Schwarzschild solution of the Robinson-Trautman class of algebraically special solutions \cite{Robinson:1962zz,exact},  for which the repeated principal null direction has vanishing rotation. The second type grows exponentially with $u$ and hence diverges on the future horizon. This corresponds to the linearization about the Schwarzschild solution of the class of algebraically special solutions for which the repeated principal null direction has non-vanishing rotation.\footnote{\label{CNtimereversal} Two more types of algebraically special perturbation are related to these by the Schwarzschild time-reversal isometry. This gives solutions which vanish on the past horizon and grow exponentially with the advanced time coordinate $v$, and solutions which diverge on the past horizon and decay exponentially with $v$.}

We will consider the $d$-dimensional generalized Schwarzschild solution for which the $(d-2)$-sphere of the metric is replaced by an arbitrary compact Einstein manifold ${\cal K}^{d-2}$ with curvature of any sign. We also include a cosmological constant. We will exploit the results of Kodama and Ishibashi \cite{Kodama:2003jz,Kodama:2000fa}, who showed that linearized metric perturbations can be decomposed into scalar, vector and tensor types on ${\cal K}^{d-2}$. For each type there is a ``master equation", a wave equation for a single scalar quantity. This can be reduced to a wave equation in $2d$ by expansion in harmonics on ${\cal K}^{d-2}$. For each type of perturbation, we show that imposing the algebraically special condition gives an additional equation so the problem reduces to solving this equation simultaneously with the master equation. Our results are as follows.

We find that algebraically special perturbations arise in the tensor sector if, and only if, there exist infinitesimal traceless perturbations of the metric on ${\cal K}^{d-2}$ that preserve the Einstein condition. 

For $d=4$, there are two infinite classes of algebraically special perturbations: one of vector type and one of scalar type. These correspond to the perturbations discovered by Couch and Newman, generalized to allow for planar or hyperbolic symmetry and a cosmological constant. 

For $d>4$, we find that algebraically special vector and scalar perturbations are much more restricted then for $d=4$. In the vector sector, algebraically special perturbations arise only when ${\cal K}^{d-2}$ has non-negative curvature and admits an isometry, in which case the perturbation corresponds to adding angular, or linear, momentum in the direction of the isometry. For ${\cal K}^{d-2}=S^{d-2}$ this corresponds to the Myers-Perry solution linearized around the Schwarzschild solution. Finally, the only algebraically special perturbation in the scalar sector corresponds to a variation of the mass of the Schwarzschild solution. 

Our result shows that $d>4$ algebraically special perturbations are stationary. We have obtained essentially the same class of perturbations as was found in Ref. \cite{Kodama:2004kz}, which determined all {\it stationary} perturbations of Schwarzschild that are regular at the horizon and vanish at infinity, for the case in which ${\cal K}^{d-2}$ is a space of constant curvature.

In summary, the only algebraically special perturbations of the higher-dimensional Schwarzschild solution correspond to simple variations of parameters in known solutions. In contrast with $d=4$, there are no time-dependent algebraically special perturbations. From the perspective of looking for new algebraically special solutions, this is disappointing. However, it is encouraging from the point of view of using the quantity $\delta \Omega_{ij}$ to study linearized metric perturbations since our result implies that this quantity determines the perturbation up to the addition of a ``variation of parameters'' perturbation. This is better than in four dimensions, where boundary conditions at the horizon and infinity are required to eliminate the Couch-Newman algebraically special perturbations \cite{wald}. 

A loop-hole in our result (and that of Couch and Newman) is that we have assumed that the perturbation is smooth on ${\cal K}^{d-2}$ to perform the scalar/vector/tensor decomposition and expansion in harmonics. So we miss algebraically special perturbations which are not smooth on ${\cal K}^{d-2}$. This includes the perturbation corresponding to turning on NUT charge or, in 4d, acceleration (i.e. the C-metric)  \cite{wald}. It turns out that the former can be studied using our approach by considering singular eigenfunctions of the Laplacian on ${\cal K}^{d-2}$. However, the singular perturbation corresponding to the 4d C-metric does not arise from a singular eigenfunction of the Laplacian on $S^2$. See Ref. \cite{Kodama:2008wf} for a discussion of higher dimensional analogues of this perturbation. 

This paper is organized as follows. Section \ref{sec:AlgSpecial} starts with the definition of algebraically special perturbations. Appendix  \ref{sec:algSpecialPert} describes the formalism used to derive the condition for an algebraically special perturbation in a type D Einstein background. In Section \ref{sec:KI} we restrict our analysis to the generalized Schwarzschild black hole solutions. We find the solutions of the algebraically special condition in these backgrounds that also solve the Kodama-Ishisbashi master equation, for the sensor, vector and scalar sector of perturbations. A final discussion of our conclusions is given in Section \ref{Conc}.

%%%%%%%%%%%%%%%%%%%%%%%%%%%%%%%%%%%%%%%%%%%%%%
\section{Algebraically special perturbations\label{sec:AlgSpecial}}
%%%%%%%%%%%%%%%%%%%%%%%%%%%%%%%%%%%%%%%%%%%%%%

As discussed in the Introduction, there are various inequivalent definitions of ``algebraically special" in higher dimensions \cite{desmet,cmpp,taghavi,Durkee:2010xq}. We will see in this section how each of these definitions leads to the same necessary condition for a {\it perturbation} to be algebraically special. 

Introduce a null basis $ \{ \eb_a\}=\{\lb,\nb,\mb{i} \}$, with $a=0,1,\ldots, d-1$, $i=2,\ldots ,d-1$, which  obeys the orthogonality relations,
\begin{equation}\label{nullbasis}
 \lb^2=n^2=\lb \cdot \mb{i} = \nb \cdot \mb{i}=0, \qquad 
 \lb \cdot \nb = 1, \qquad 
 \mb{i} \cdot \mb{j} = \delta_{ij}.
\end{equation}
We will use Latin indices to label components of a tensor in the null basis and Greek letters for abstract indices.
The following notation is used for certain Weyl tensor components \cite{Durkee:2010xq}
\be
 \Omega_{ij} = C_{0i0j} =  \lb^\alpha \mb{i}^\beta \lb^\mu \mb{j}^\nu  C_{\alpha \beta \mu \nu}\,,
\ee
\be
 \Psi_{ijk} = C_{0ijk} = \lb^\alpha \mb{i}^\beta \mb{j}^\mu \mb{k}^\nu C_{\alpha \beta \mu \nu}\,, \qquad \Psi_i = \Psi_{jij}
\ee
\be
 \Phi^S_{ij} = C_{0(i|1|j)}, \qquad \Phi^A_{ij} = \frac{1}{2} C_{01ij}, \qquad \Phi = \Phi^S_{ii}
\ee
Following Ref. \cite{cmpp}, the null vector $\lb$ is called a Weyl Aligned Null Direction (WAND) if
\be
 \Omega_{ij} = 0\,.
\ee
It is called a multiple WAND if
\be
 \Omega_{ij} = \Psi_{ijk}= 0\,.
\ee
These definitions do not depend on how the other basis vectors are chosen \cite{cmpp}. In 4d, a WAND is the same as a principal null direction and a multiple WAND the same as a repeated principal null direction. Hence, in 4d, a spacetime is algebraically special if, and only if, it admits a multiple WAND. In $d>4$ dimensions, some references define a spacetime to be algebraically special if it admits a WAND \cite{cmpp} and others require a multiple WAND \cite{Durkee:2010xq}. The definition of Ref. \cite{taghavi} is stronger still: a solution which is algebraically special according to this definition must admit a multiple WAND and satisfy some additional conditions. 

The Myers-Perry solution admits a multiple WAND, in fact it admits two distinct multiple WANDs \cite{Frolov:2003en,Hamamoto:2006zf}, which implies that it is type D in the classification of Ref. \cite{cmpp}. It is also algebraically special according to the definition of Ref. \cite{taghavi}. Of course the same remarks apply to the Schwarzschild solution. 

To obtain a necessary condition for a perturbation to preserve the algebraically special property, consider a one-parameter family of algebraically special solutions specified by a parameter $\lambda$, such that for $\lambda=0$, the solution is the Schwarzschild solution of mass $M$. A necessary condition for the family of solutions to be algebraically special according to any of the definitions of Refs. \cite{cmpp,taghavi,Durkee:2010xq} is that there exists a WAND. We assume that this depends smoothly on $\lambda$. For $\lambda=0$ this must reduce to one of the multiple WANDs of the Schwarzschild solution (since every WAND of Schwarzschild is a multiple WAND\footnote{\label{pravda}
Any null vector can be related to one of the multiple WANDs by a ``null rotation". But one can show that the only such rotation that preserves the WAND condition is the trivial one.}). Hence, by choosing the basis vector $\ell^a$ to be this WAND, we have $\Omega_{ij}=0$ for all $\lambda$. 

Differentiating with respect to $\lambda$ and evaluating at $\lambda=0$ gives
\be
\label{algSpecPert}
 \delta \Omega_{ij} = 0\,.
\ee
This is a necessary condition for a linearized perturbation of the Schwarzschild solution to be the linearization around Schwarzschild of a larger family of solutions admitting a WAND which depends smoothly on $\lambda$. We will define an algebraically special perturbation to be a solution of this equation. The LHS of this equation has the desirable property that it is  gauge invariant under infinitesimal coordinate transformations and infinitesimal basis transformations \cite{Durkee:2010qu}. 

Our assumption that the WAND depends smoothly on $\lambda$ is highly non-trivial. For example, in 4d, {\it any} metric admits a WAND. However, one can show that, in 4d, some components of a WAND (with respect to a basis depending smoothly on $\lambda$) generically behave as $\sqrt{\lambda}$ as $\lambda \rightarrow 0$ and hence are not differentiable with respect to $\lambda$ at $\lambda=0$. 

We will now show that the WAND {\it does} depend smoothly on $\lambda$ if it satisfies the additional (basis-independent) condition $\Psi_i=0$. Introduce a null basis $\{\bar{\ell},\bar{n},\bar{m}_i \}$ which depends smoothly on $\lambda$ with the property that $\bar{\ell}(0)$ is a multiple WAND of Schwarzschild. Now expand the WAND $\ell(\lambda)$ in this basis:
$\ell^a = x \left( \bar{\ell}^a + z_i \bar{m}_i^a - (1/2) z_k z_k \bar{n}^a \right)$ where $x,z_i$ are functions of $\lambda$ with $x(0)=1, z_i(0)=0$ (and we have used the fact that $\ell^a$ is null). Without loss of generality we can set $x \equiv 1$. Then $\ell^a$ is related to $\bar{\ell}^a$ by a null rotation about $\bar{n}^a$ with parameters $z_i$. Using the transformation properties of $\Psi_i$ under a null rotation  \cite{Durkee:2010xq} we have
\be
\label{Psinullrot}
 \Psi_i =  \bar{\Psi}_i - \bar{\Phi} z_i + 3 \bar{\Phi}^A_{ij} z_j - \bar{\Phi}_{ij}^S z_j + {\cal O}(z^2) \,,
\ee
where the overbar refers to the smooth basis. If we assume that our WAND obeys $\Psi_i=0$ then the implicit function theorem implies that $z_i$ must depend smoothly on $\lambda$ in a neighbourhood of $\lambda=0$. We just have to check that $\det \partial_{z_j} \Psi_i \ne 0$ in a neighbourhood of $\lambda=0,z_i=0$. This is indeed true because, at $\lambda=0,z_i=0$ we have
\be
 \partial_{z_j} \Psi_i = -\bar{\Phi} \delta_{ij} + 3 \bar{\Phi}^A_{ij} - \bar{\Phi}^S_{ij} = -\frac{d-1}{d-2}\, \bar{\Phi} \,\delta_{ij}\,,
\ee
 where on the RHS we made use of the relations $\bar{\Phi}^S_{ij} = (\bar{\Phi} /(d-2)) \delta_{ij}$ and $\bar{\Phi}^A_{ij}=0$ which are satisfied in the Schwarzschild spacetime. The determinant of the RHS is non-zero (because $\bar{\Phi} \ne 0$ in the Schwarzschild solution) and so the result follows. 

We have shown that if our family of solutions admits a WAND satisfying the extra conditions $\Psi_i=0$ then the WAND depends smoothly on $\lambda$ and hence the resulting linearized perturbation of Schwarzschild will obey (\ref{algSpecPert}). There is a partial converse to this. Assume that we have a linearized perturbation of Schwarzschild which satisfies (\ref{algSpecPert}). Then equation (\ref{Psinullrot}) linearized in $z_i$ shows that we can perform an infinitesimal null rotation to set $\delta \Psi_i=0$, which is the linearized version of $\Psi_i=0$. 

In 4d, the condition $\Psi_i=0$ is equivalent to $\Psi_{ijk}=0$. So in 4d, existence of a WAND with $\Psi_i=0$ is the standard condition for the family of solutions to be algebraically special. In $d>4$ dimensions, $\Psi_i=0$ is weaker than the multiple WAND condition $\Psi_{ijk}=0$. In the classification of Ref. \cite{cmpp}, existence of a WAND with $\Psi_i=0$ is the condition for a solution to be type I(a) or more special. Hence any 1-parameter family of solutions that is type I(a) or more special will give a solution of (\ref{algSpecPert}) when linearized around the Schwarzschild solution. 

So far we have discussed the notions of algebraically special arising in Refs. \cite{cmpp,taghavi,Durkee:2010xq}. We will now discuss the definition of Ref. \cite{desmet}. The latter definition is restricted to $d=5$ and is based on a spinorial classification of the Weyl tensor. In this definition, the basic object is the Weyl spinor $P_{ABCD}=(C\Gamma^{ab})_{AB} (C\Gamma^{cd})_{CD} C_{abcd}$, where $C$ denotes the charge-conjugation matrix and spinor indices $A,B,C,D$ take values from $1$ to $4$. $P_{ABCD}$ is totally symmetric \cite{desmet} so contracting with a Dirac spinor $\psi^A$ gives $P(\psi) \equiv P_{ABCD} \psi^A \psi^B \psi^C \psi^D$, a homogeneous polynomial of degree $4$ depending on the $4$ complex components of $\psi^A$. Ref. \cite{desmet} classified the Weyl tensor according to whether, and how, this polynomial factorizes into polynomials of lower degree. An algebraically special solution is one for which the polynomial factorizes.

The Myers-Perry and Schwarzschild solutions are algebraically special: $P(\psi)$ is the square of a quadratic polynomial \cite{DeSmet:2003kt}. This means that these solutions are type $\underline{\bf 22}$ in the classification of Ref. \cite{desmet}. Now, as above, consider a family of solutions that are algebraically special according to this definition, and reduce to Schwarzschild when $\lambda=0$. Ref. \cite{godazgar} showed that the possible factorizations of $P(\psi)$ are restricted by a reality condition. The only allowed factorization that can reduce to the square of a quadratic polynomial when $\lambda=0$ corresponds to $P(\psi)$ being the product of two quadratic factors. If these are distinct then this corresponds to type ${\bf 22}$ in the classification of Ref. \cite{desmet}. 

The coefficients of $P(\psi)$ depend smoothly on $\lambda$. However, we cannot expect the coefficients of the quadratic factors to be differentiable with respect to $\lambda$ at $\lambda=0$. This is because the quadratic factors become coincident at $\lambda=0$. For example, write $\psi^A=(w,x,y,z)$ and consider $P(\psi)=(w^2+\sqrt{\lambda} (x^2+y^2+z^2)) (w^2 - \sqrt{\lambda} (x^2+y^2+z^2))$: the coefficients of $P$ are smooth at $\lambda=0$ but the coefficients of the quadratic factors are not. However, if our family is of type $\underline{\bf 22}$ for all $\lambda$ then $P(\psi)=Q(\psi)^2$ for some quadratic polynomial $Q$ and the coefficients of $Q$ {\it will} be smooth functions of $\lambda$, even at $\lambda=0$. To see this, pick $\psi_0$ such that $P(\psi_0) \ne 0$ at $\lambda=0$. In a neighbourhood of $\psi=\psi_0, \lambda=0$ we have $P(\psi) \ne 0$ and then $Q(\psi) = \sqrt{P(\psi)}$ implies that the coefficients in $Q$ depend smoothly on $\lambda$ at $\lambda=0$. 

The Weyl tensor of a type $\underline{\bf 22}$ solution can be written as an expression quadratic in an antisymmetric tensor $A_{\mu\nu}$ \cite{godazgar} which is constructed linearly from $Q$ and hence depends smoothly on $\lambda$. If one introduces a null basis as above then one has \cite{godazgar}
\be
 \Omega_{ij} = A_{0k} A_{0k} \delta_{ij} - 3 A_{0(i} A_{|0|j)} \,.
\ee
Linearizing around the Schwarzschild background gives
\be
\label{deltaOmega2}
 \delta \Omega_{ij} = 2 A_{0k}  \delta A_{0k}  \delta_{ij} - 6 A_{0(i} \delta A_{|0|j)}\,,
\ee
where $A_{0i}$ is evaluated in the Schwarzschild geometry. Let us now choose our null basis as before, i.e., so that $\ell^a$ is a (multiple) WAND of Schwarzschild when $\lambda=0$. This implies that $A_{0i}=0$ \cite{godazgar} so (\ref{deltaOmega2}) reduces to (\ref{algSpecPert}). Hence (\ref{algSpecPert}) is a necessary condition for a perturbation to correspond to the linearization about Schwarzschild of a family of type $\underline{\bf 22}$ solutions.\footnote{This result generalizes to a family of type ${\bf 22}$ if one {\it assumes} that the quadratic polynomials depend smoothly on $\lambda$. However, as discussed in the text, this assumption is not expected to be generally valid.} 

In summary, we have shown that (\ref{algSpecPert}) is a necessary condition for a linearized perturbation of the Schwarzschild solution to arise from a 1-parameter family of solutions which is type I(a) or more special in the classification of Ref. \cite{cmpp} (as will be the case if it is algebraically special according to the definitions of Refs. \cite{taghavi,Durkee:2010xq}) or, in 5d, is of type $\underline{\bf 22}$ in the classification of Ref. \cite{desmet}. 

As mentioned in the Introduction, equation (\ref{algSpecPert}) arises also in the study of general perturbations of algebraically special solutions. For perturbations of a solution admitting a multiple WAND (e.g. Myers-Perry), $\delta \Omega_{ij}$ is gauge invariant under infinitesimal coordinate transformations and infinitesimal basis transformations \cite{Durkee:2010qu}. This makes $\delta \Omega_{ij}$ a natural quantity to consider when studying perturbations of such a solution. Furthermore, since $\delta \Omega_{ij}$ is a $(d-2) \times (d-2)$ traceless symmetric matrix, it has the same number of degrees of freedom as the gravitational field and so it seems likely that the metric perturbation could be reconstructed from knowledge of $\delta \Omega_{ij}$. However, $\delta \Omega_{ij}$ determines the metric perturbation only up to addition of a solution of (\ref{algSpecPert}). Hence it is desirable to determine all solutions of (\ref{algSpecPert}) in order to determine the extent to which $\delta \Omega_{ij}$ uniquely characterizes the metric perturbation. 

Our definition of an algebraically special perturbation as a solution of (\ref{algSpecPert}) requires choosing our basis vector $\ell^a$ to reduce to one of the multiple WANDs of Schwarzschild when $\lambda=0$. But there are two such multiple WANDs. A perturbation which is algebraically special with respect to one choice generically will not be special with respect to the other choice. However, the two multiple WANDs of Schwarzschild are related by the time-reversal isometry. Hence by applying this isometry to our solutions we will ensure that we do not miss any algebraically special perturbations. For $d=4$ the result of doing this is described in footnote \ref{CNtimereversal}. 

The calculation of $\delta \Omega_{ij}$ for a general metric perturbation of a type D Einstein spacetime is explained in Appendix \ref{sec:algSpecialPert}.

%%%%%%%%%%%%%%%%%%%%%%%%%%%%
\section{Kodama-Ishibashi decomposition} \label{sec:KI} 
%%%%%%%%%%%%%%%%%%%%%%%%%%%%

%%%%%%%%%%%%%%
\subsection{Introduction}

Our background geometry is the generalized Schwarzschild solution in $d$ dimensions. In ingoing Eddington-Finkelstein coordinates, this is
\begin{equation}
 ds^2 =-f(r)dv^2+2 dv dr + r^2 \gamma_{ij} dx^i dx^j   \qquad \hbox{with}\qquad  f(r) = K  -\lambda_c \,r^2 - \frac{r_m^{d-3}}{r^{d-3}}\,, 
\label{gSch}
\end{equation}
Here  $\gamma_{ij}$ is the metric on a $(d-2)$-dimensional compact Einstein manifold ${\cal K}^{d-2}$ with Ricci tensor $K(d-3)\gamma_{ij}$ where $K\in \{ 0,\pm 1 \}$. $r_m$ is a mass parameter while $\lambda_c$ is the cosmological constant parameter appearing in the Einstein equation $R_{\mu\nu} = (d-1) \lambda_c\, g_{\mu\nu}$. 

The standard Schwarzschild(-de Sitter) solution has $K=1$ and ${\cal K}^{d-2} = S^{d-2}$. If $\lambda_c<0$, the solutions with $K=0$ and $K=-1$ are also regular black holes. These include the planar and hyperbolic AdS-Schwarzschild black holes, with $T^{d-2}$ and (compactified) $H^{d-2}$ horizon topology, respectively.

We will sometimes write the metric in the form
\be
ds^2 = g_{AB}dx^A dx^B+r^2\gamma_{ij} dx^i dx^j ,
\ee
where $g_{AB}$ is the Lorentzian metric of the two-dimensional orbit spacetime.

In our computations we will use the null basis 
\begin{equation}
\lb= dv\,,\quad \nb= dr -\frac{1}{2}\,f dv\,,\quad \mb{i}=r\hat{e}_i \,,
\label{Schw:nullbasis}
\end{equation}  
where $\hat{e}_i$ are a vielbein for the metric $\gamma_{ij}$. Note that $\ell$ and $n$ are the multiple WANDs of Schwarzschild.

We can decompose perturbations according to how they transform under diffeomorphisms of ${\cal K}^{d-2}$. This decomposition was worked out by Kodama and Ishibashi \cite{Kodama:2003jz}. An arbitrary metric perturbation $h_{\mu\nu}$ can be decomposed into perturbations of scalar, transverse vector, and transverse traceless tensor types on ${\cal K}^{d-2}$. Note that in performing this decomposition we exploit our assumption that ${\cal K}^{d-2}$ is compact, and assume that $h_{\mu\nu}$ is regular on ${\cal K}^{d-2}$. 

Kodama and Ishibashi showed that each type of perturbation can be encoded in a gauge invariant scalar quantity which satisfies a ``master equation". Expanding in harmonics on ${\cal K}^{d-2}$ (again assuming compactness and regularity), this can be reduced to a wave equation in the $2d$ orbit space with metric $g_{AB}$. We will solve this equation simultaneously with the equation arising from the algebraically special condition.

The latter condition  $\delta \Omega_{ij}=0$ for a general metric perturbation of a type D Einstein spacetime is given in \eqref{dOmegaij} of Appendix \ref{sec:algSpecialPert}.
In the Schwarzschild background it simplifies considerably.
With the harmonic decomposition we can write it in terms of the gauge invariant master variable and it reduces to the vanishing of a product of two factors. One factor depends only on the harmonic and its derivatives. The other contribution depends only on the 2-dimensional orbit spacetime coordinates $v,r$. Therefore, to have $\delta \Omega_{ij}=0$, one of these two factors must vanish. Typically we will find that it is the orbit spacetime factor that constrains our search, although we will also encounter special cases where the algebraically special condition is automatically obeyed because the contribution from the harmonic vanishes (these cases describe the perturbations that just shift the mass and angular momentum).

%%%%%%%%%%%%%%
\subsection{Tensor perturbations} \label{sec:KItensor} 
%%%%%%%%%%%%%%

Tensor-type perturbations have the form \cite{Gibbons:2002pq,Kodama:2003kk}
%\footnote{Note that, for Einstein base spaces, tensor perturbations are expanded in terms of the eigentensors of the Lichnerowicz operator $\Delta_L$ \cite{Gibbons:2002pq,Kodama:2003kk}. (For maximally symmetric base spaces, the expansion is in terms of the tensor harmonics of the Laplacian operator $-D^2$).}
\begin{eqnarray}
h_{AB}=0, \quad  h_{Ai}=0,\quad   h_{ij}=2r^2 H_T \tensor_{ij} \,,  
\label{pert:tensor}
\end{eqnarray} 
where $H_T=H_T(v,r)$, and  $\tensor_{ij}(x^k)$ are  eigentensors obeying 
\be
\left( \Delta_L -\lambda_L \right )\tensor_{ij}=0,\qquad \hbox{with} \qquad  \Delta_L\tensor_{ij}= -D^2\tensor_{ij} -2 R_{ikjl}\tensor^{kl}+ 2(d-3)K \tensor_{ij},
\ee
where  $\Delta_L$ is the Lichnerowicz operator, $\lambda_L$ is the associated eigenvalue, $D$ is the derivative defined by the metric $\gamma_{ij}$ of the Einstein base space ${\cal K}^{d-2}$, and $R_{ikjl}$ is the associated Riemann tensor  \cite{Gibbons:2002pq,Kodama:2003kk}.

The Kodama-Ishibashi gauge invariant master variable for a tensorial perturbation is 
\be
\Phi_T = r^{(d-2)/2}H_T.
\ee
It obeys the master equation  \cite{Gibbons:2002pq,Kodama:2003kk}
\begin{equation} 
 \left(  \Box_2 - \frac{V_T}{f} \right) \Phi_T =0 \,, \qquad V_T = \frac{f}{r^2}  
         \left[   \frac{d(d-2)}{4}f+\frac{(d-1)(d-2)r_m^{d-3}}{2 r^{d-3}}+\lambda_L-(3d-8)K \right],
\label{master:tensor} 
\end{equation}  
where $\Box_2$ is the d'Alembertian operator in the 2-dimensional orbit spacetime with metric $g_{AB}$. 

Written in terms of the master variable we find that the algebraically special condition is 
\be
\delta \Omega _{ij}= 0  \quad \Leftrightarrow  \quad \partial_r^2 \left(r^{-\frac{d-2}{2}}\Phi_T \right)+\frac{2}{r}\partial _r \left(r^{-\frac{d-2}{2}}\Phi_T \right)=0 \,, 
\ee
with solution
 \be
   \Phi_T=r^{\frac{d-2}{2}}\left(\frac{H_1(v)}{r}+H_2(v)\right)
\ee
for arbitrary functions $H_1$ and $H_2$ of the Eddington-Finkelstein coordinate $v$. Inserting this expression for $\Phi_T$ into the master equation \eqref{master:tensor}  we find that the latter reduces to a polynomial in $r$: 
\begin{eqnarray}\label{tensor1}
&&(d-2)\left[H_2'(v)+\lambda_c \,H_1(v)\right]r^{d-1}+ \left[(d-4) H_1'(v)-\left(  \lambda_L-2(d-3)K \right) H_2(v)\right] r^{d-2}\nonumber\\
&&\hspace{1cm} - \left[\lambda_L-(d-2) K\right] H_1(v)r^{d-3}-r_m^{d-3} H_1(v)=0\,.
\end{eqnarray}
For $\lambda_L \neq 2(d-3)K$, only the trivial solution $H_1(v)=0=H_2(v)$ satisfies this equation (we assume $r_m \ne 0$). For $\lambda_L= 2(d-3)K$, the general solution is $H_1=0$, $H_2={\rm constant}$. 
In summary, the algebraically special tensorial perturbations are given by:
\begin{equation}
\lambda_L= 2(d-3)K, \qquad \Phi_T=H_2 \,r^{\frac{d-2}{2}} \,.
\end{equation}
Tensor harmonics with $\lambda_L= 2(d-3)K$ are linearized perturbations of the metric on ${\cal K}^{d-2}$ which preserve the Einstein condition (and the volume of ${\cal K}^{d-2}$). So algebraically special tensor perturbations exist if, and only if, ${\cal K}^{d-2}$ admits such perturbations. If ${\cal K}^{d-2}$ is a space of constant curvature then such perturbations exist only if either (i) $K=0$ or (ii) $d=4$ and $K=-1$ (see e.g. Ref. \cite{Kodama:2004kz}).

%%%%%%%%%%%%%%
\subsection{Vector perturbations.} \label{sec:KIvector} 
%%%%%%%%%%%%%%

Vector perturbations are constructed out of vector harmonics $\vector_i(x^k)$  \cite{Kodama:2003jz}
\begin{equation}\label{pert:vector}
  h_{AB}=0 \,, \quad 
  h_{Ai}=r f_A \vector_i ,\quad 
  h _{ij} = -\frac{2}{k_V}r^2 H_T  D_{(i}\vector_{j)} \,, 
\end{equation} 
where $f_A$ and $H_T$ are functions of $\{x^A\}=\{v,r\}$, and $\vector_i$ denotes a transverse vector harmonic on ${\cal K}^{d-2}$: 
\be
D_i \vector^i=0, \qquad \left (D^2 + k_V^2 \right) \vector_i=0
\ee
If ${\cal K}^{d-2} = S^{d-2}$ then the eigenvalues are
\be
\label{veceval}
 k_V^2=l(l + d-3) -1, \qquad l=1,2,\ldots
\ee
Harmonics with $k_V^2=(d-3)K$ ($l=1$ above) are special: they are Killing vectors on (compact) ${\cal K}^{d-2}$, occuring only for $K=0,1$. The cases $k_V^2\neq (d-3)K$ and $k_V^2=(d-3)K$ are described by different gauge invariant quantities and thus we will discuss them separately in the following two subsections.

Vectorial gauge transformations, $h_{\mu\nu}\to h_{\mu\nu}+ {\cal L}_\xi g_{\mu\nu}$,  are generated by an infinitesimal gauge vector $\xi$ with harmonic decomposition  \cite{Kodama:2003jz}, 
\be
\xi=r \,L(x^A)\, \vector_{i} \,dx^i .\label{VectorGauge}
\ee

%%%%%%%%%%
\subsubsection{Modes with $\bm{k_V^2 \neq (d-3)K}$}
%%%%%%%%%%

If ${\cal K}^{d-2} = S^{d-2}$ then this corresponds to $l>1$.

A master variable $\Phi_V$ can be constructed in terms of a variable $F_{A}$ which is a  gauge invariant combination of $f_A$ and $H_T$, 
\begin{eqnarray} \label{FfunctionPhi}
   F_A \equiv f_A + \frac{r}{k_V} D_A H_T 
       = r^{-(d-3)} \epsilon_{AB} D^B \left(r^{(d-2)/2} \Phi_V\right),
\end{eqnarray}   
where $\epsilon_{AB}$ denotes the volume form on the $2d$ orbit spacetime. The quantity $\Phi_V$ obeys the Kodama-Ishibashi master equation \cite{Kodama:2003jz}
\begin{equation} 
 \left(  \Box_2 - \frac{V_V}{f} \right) \Phi_V = 0 \,, \qquad 
 V_V = \frac{f}{r^2}    \left[  k_V^2 -(d-3)K + \frac{d(d-2)}{4}f  
                 - \frac{d-2}{2} r f'
            \right] .
\label{master:vector} 
\end{equation}  
We find that the algebraically special condition is
\be
 \delta \Omega _{ij}= 0   \quad \Leftrightarrow  \quad \partial _r^2\Phi _V+\frac{2 }{r}\partial _r\Phi _V-\frac{(d-2)(d-4)}{4}\frac{ \Phi _V }{r^2}=0 \,,
\ee
with solution
\be  
\Phi _V=\frac{C_0(v)}{r^{(d-2)/2}}+C_1(v)r^{(d-4)/2}
\ee
for arbitrary functions $C_{0,1}(v)$ of the ingoing Eddington-Finkelstein coordinate $v$. 
Inserting this into the master equation \eqref{master:vector} gives a polynomial in $r$: 
\begin{eqnarray}\label{masterAnyd:vector}
&&\hspace{-1.5cm} -(d-4) C_1'(v) r^{d-2}+\left(k_V^2+(d-3)K \right)C_1(v) r^{d-3} +(d-2)C_0'(v) r \nonumber\\
&&\hspace{0.5cm} +\left[\left(k_V^2-(d-3) K\right) C_0(v)-(d-1)(d-3) r_m^{d-3}C_1(v)\right]=0
\,.
\end{eqnarray}
For $d\geq 5$ this polynomial involves four distinct powers of $r$. However,  $d=4$ is special because then the polynomial has degree 1. For this reason we analyze  the   $d=4$ and $d \ge 5$ cases separately.

\vskip 0.2cm
\noindent{\bf  i) Case $\bm{d=4}$}
\vskip 0.1cm

For $d=4$, equation \eqref{masterAnyd:vector} reduces to
\begin{equation}\label{masterd4:vector}
-\left[\left(K-k_V^2\right) C_0(v)+3 r_m C_1(v)\right]+r \left[\left(K+k_V^2\right) C_1(v)+2 C_0'(v)\right]=0
\,,
\end{equation}
which can be satisfied only if the two coefficients of the polynomial vanish independently. 
This gives
\begin{equation}\label{S1masterd4:vector}
C_0(v)=-\frac{3 r_m C_1(v)}{K-k_V^2}, \qquad C_1(v)=A_0\left(K-k_V^2\right) e^{-(k_V^4-K^2)v/(6 r_m)} \,,
\end{equation}
for an arbitrary constant $A_0$ (recall that modes with $k_V^2 = (d-3)K=K$ are excluded from the analysis of this subsection).
The associated metric perturbations can be reconstructed using (\ref{FfunctionPhi}) once a gauge is chosen. This map simplifies if we take advantage of the gauge transformation described by \eqref{VectorGauge} to pick the 
 gauge $H_T=0$. The perturbation just obtained is then 
 \begin{equation}\label{Sh1masterd4:vector}
f_v=- f_r\left(f(r)+\frac{ \left(K+k_V^2\right)}{6r_m}\left[\left(K-k_V^2\right) r-3 r_m\right]\right),\quad f_r=-\frac{ A_0\left(K-k_V^2\right)}{r}\,e^{-(k_V^4-K^2)v/(6 r_m)}.
\end{equation}
Recall that $f(r)$ is the function appearing in the Schwarzschild metric.

We have found an infinite class of algebraically special perturbations labelled by the eigenvalue $k_V^2$. In general, these perturbations decay exponentially with $v$ and blow up on the past horizon.\footnote{\label{timereversal} As discussed at the end of section \ref{sec:AlgSpecial}, one obtains another solution by applying the time reversal isometry. This solution grows exponentially with the retarded time coordinate $u$ and diverges on the future horizon.} For $K=1$ and $\Lambda=0$ these perturbations were first identified by Couch and Newman \cite{Couch:1973zc}. They correspond to the linearization about the Schwarzschild solution of the class of algebraically special solutions for which the repeated principal null direction has non-vanishing rotating. 

Interestingly, the above perturbation \eqref{S1masterd4:vector} becomes time independent when $k_V^2=- K$ (recall $k_V^2=K$ is excluded from this section). Indeed,  \eqref{S1masterd4:vector} and  \eqref{Sh1masterd4:vector}  reduce to
\begin{eqnarray}\label{NutS1masterd4:vector}
&& C_0(v)=-\frac{3 r_m C_N}{2K}, \qquad C_1(v)=C_N\,;\nonumber \\
&& f_v=\frac{C_N f}{r}\,,\quad f_r=-\frac{C_N}{r}\,,\quad H_T=0\,,
\end{eqnarray}
where we redefined the arbitrary constant of the problem as $C_N \equiv 2 K A_0$. The interpretation of this perturbation has been discussed in Ref. \cite{Kodama:2004kz}. 

If $K=1$ then $k_V^2 = -K = -1$ cannot correspond to a regular vector harmonic on $S^2$. For $K=-1$, we will show at the end of this section that if the harmonic is regular then the perturbation is locally pure gauge. However one can find non-trivial {\it singular} harmonics with this eigenvalue (which corresponds to setting $l=0$ in (\ref{veceval})) \cite{Kodama:2004kz}. To see this, let $(\theta,\phi)$ be spherical polars on $S^2$. An example of a vector harmonic on $S^2$ with $k_V^2 = -1$ is $\vector = \cos \theta  d\phi$. This is singular at $\theta=0,\pi$.

This perturbation adds NUT charge, proportional to the NUT parameter $N\equiv C_N/2$, to the Schwarzschild black hole \cite{Kodama:2004kz}. To confirm this statement, recall that  the (A)dS$-$Taub-NUT solution is described by the line element  \cite{exact}\begin{eqnarray}
\begin{aligned}
 &ds^2 =-g(r)\left(dt+2 N \cos\theta d\phi \right)^2+\frac{dr^2}{g(r)}+\left(r^2+N^2\right)d\Omega_2^2\,,
\nonumber\\
 & \hbox{with} \quad g(r)=\frac{ r^2-2M r- N^2-\lambda_c\left(r^4+ 6N^2r^2-3 N^4\right)}{r^2+N^2}\,.
\end{aligned}
\label{NutBH}
\end{eqnarray}
Converting to Eddington-Finkelstein coordinates and linearizing in $N$ reproduces the perturbation just discussed. A similar result holds for $K=-1$. 

\vskip 0.2cm
\noindent{\bf  ii) Case $\bm{d>5}$: addition of NUT charge}
\vskip 0.1cm

For $d\geq 5$, there are no time-dependent solutions of equation \eqref{masterAnyd:vector}. The only solution is
\begin{equation}\label{NutS1masterd5:vector}
k_V^2=-(d-3)K, \qquad C_0(v)=-\frac{(d-1)(d-3)  r_m^{d-3}C_N}{2(d-3) K} ,\qquad C_1(v)=C_N\,,
\end{equation}
for arbitrary constant $C_N$ and $K=\pm 1$. This solution is the natural higher dimensional generalization of \eqref{NutS1masterd4:vector}. In the gauge $H_T=0$, the associated metric perturbation is $f_v=C_N f/r$, $f_r=-C_N/r$. Converting to Schwarzschild coordinates this gives
\be
\label{TNpert}
h_{\mu\nu} dx^\mu dx^\nu = 2 C_N f V_i dt dx^i
\ee
For even $d$, taking ${\cal K}^{d-2}$ to be a positive K\"ahler-Einstein space, this perturbation corresponds to the linearization of the higher-dimensional Taub-NUT solution of Ref. \cite{Bais:1984xb}. For general $d>5$, taking ${\cal K}^{d-2}$ to be a product ${\cal K}_1^{2n} \times {\cal K}_2^{d-2-2n}$ where ${\cal K}_1^{2n}$ is K\"ahler-Einstein and ${\cal K}_2^{d-2-2n}$ is Einstein, this perturbation corresponds to the linearization of the higher-dimensional Taub-NUT solutions of Refs. \cite{Mann:2003zh,Lu:2004ya}. For $d=6$, the algebraic type of some of these solutions was discussed in Ref. \cite{Ortaggio:2012cp}, where they were shown to be type D in the classification of Ref. \cite{cmpp}. 

This perturbation does not correspond to a regular harmonic on ${\cal K}^{d-2}$. This is obvious for $K=1$ because it has $k_V^2<0$ and hence must be singular on ${\cal K}^{d-2}$.  For $K=-1$, note that $k_V^2 = - (d-3)K$ implies that the 1-form $\vector_i$ is harmonic with respect to the Hodge-de Rham Laplacian. If $\vector_i$ is assumed regular on compact ${\cal K}^{d-2}$, this implies that $\vector_i$ is closed so locally on ${\cal K}^{d-2}$ we can write $\vector_i = \partial_i \alpha$ for some function $\alpha$. But then from (\ref{TNpert}) one sees that the perturbation can be gauged away by a shift $t \rightarrow t+C_N \alpha$. So if the perturbation is regular on ${\cal K}^{d-2}$ then it is locally trivial. This is not the case for the Taub-NUT perturbation just discussed so it does not correspond to a regular harmonic on ${\cal K}^{d-2}$.

\vskip 0.2cm
%%%%%%%%%%
\subsubsection{Vector modes with $\bm{k_V^2 = (d-3)K}$. Addition of angular momentum}
%%%%%%%%%%

Considering only regular harmonics on ${\cal K}^{d-2}$, we must have $K=0,1$ and these harmonics are Killing vector fields so there is no $H_T$ contribution in \eqref{pert:vector}. The unique gauge invariant variable in this case is $F_{AB}$. Its definition, the master equation it must obey, and the respective solution are  \cite{Kodama:2003jz}
\begin{equation}
F_{AB}=2 r  D_{[A}\left( r^{-1}f_{B]}\right); \qquad  D^B\left( r^{d-1}F_{AB}\right)=0 \quad \Leftrightarrow \quad F_{AB}= \epsilon_{AB} \,\frac{C_J}{r^{d-1}}\,,
\label{master:vectorSpecial} 
\end{equation} 
where $D$ is the covariant derivative in the 2-dimensional orbit spacetime with metric $g_{AB}$, $\{f_A\}=\{f_v,f_r\}$ is defined in \eqref{pert:vector}, $\epsilon_{AB}$ denotes the anti-symmetric tensor in the orbit spacetime, and $C_J$ is an arbitrary integration constant.

Recall that after the harmonic expansion, $\delta \Omega_{ij}$ is given by the product of an orbit space and a base space contributions.
For the regular perturbations with $k_V^2 = (d-3)K$, the algebraically special condition $\delta \Omega_{ij}=0$ is trivially obeyed because its base space factor vanishes.

By a choice of gauge, the algebraically special perturbation \eqref{master:vectorSpecial}  can be written in the form  \eqref{pert:vector} with (no $H_T$ contribution)
\be
 f_v = \frac{C_J}{(d-1) r^{d-2}}, \qquad f_r=0.
\ee
For $K=1$, this perturbation corresponds to adding angular momentum. For ${\cal K}^{d-2} = S^{d-2}$ it arises from the Kerr-Myers-Perry-(A)dS black hole \cite{myersperry,Carter:1968ks,Hawking:1998kw,Gibbons:2004uw} linearized for small angular momentum parameter(s) \cite{Kodama:2003jz}. For $K=0$, the only regular Killing vectors on compact ${\cal K}^{d-2}$ are translations (i.e. covariantly constant) and this perturbation corresponds to a boost along a translationally invariant direction.

%%%%%%%%%%%%%%%%%%%%%%%
\subsection{Scalar perturbations.} \label{sec:Kodama:2000fcalar} 
%%%%%%%%%%%%%%%%%%%

Scalar perturbations are given by 
\begin{eqnarray}\label{pert:scalar}
 h_{AB}= f_{AB} \scalar, \quad  
 h_{Ai}= rf_A \scalar_i  , \quad 
 h_{ij}= 2r^2 \left( H_L\gamma_{ij}\scalar + H_T \scalar_{ij} \right) \,,
\end{eqnarray}
with $f_{AB},f_A,H_T,H_L$ functions of $\{x^A\}=\{v,r\}$, $\gamma_{ij}$ is the base space ${\cal K}^{d-2}$ metric. $\scalar$ is  the scalar harmonic which satisfies the eigenvalue equation 
\be\label{EOMscalarH}
(D^2 + k_S^2 ) \scalar=0.
\ee
Out of this scalar harmonic we can construct a  scalar-type  vector harmonic  $\scalar_i$ and a  traceless scalar-type tensor harmonic  $\scalar_{ij}$ as (for $k_S^2 \ne 0$)
\begin{equation}
  \scalar_i = -\frac{1}{k_S}  D_i \scalar \,, \qquad 
  \scalar_{ij} = \frac{1}{k_S^2}  D_i D_j\scalar  
                 + \frac{1}{d-2}\gamma_{ij}\scalar \,.      
\end{equation}
For ${\cal K}^{d-2} = S^{d-2}$ the eigenvalues of \eqref{EOMscalarH} are
\be
 k_S^2 = l(l+d-3), \qquad l=0,1,\ldots
\ee
We must distinguish two special cases  \cite{Kodama:2003jz}. In the first special case $k_S^2=0$, i.e., constant $\scalar$, we define $\scalar_{ij}=\scalar_i=0$. These modes preserve the symmetry of the background solution. Birkhoff's theorem implies that the only such solution arises from variation of the mass parameter. 

The second special case $k_S^2 = (d-2)K \ne 0$ is possible only for $K=1$. In this case, $\scalar_{ij}=0$ so $\scalar_i$ is a conformal Killing vector (but not a Killing vector) on ${\cal K}^{d-2}$. This happens only for ${\cal K}^{d-2}=S^{d-2}$ \cite{yano}, for which it corresponds to $l=1$ perturbations. For $d=4$, such perturbations are known to be pure gauge. We will show below that the same is true for $d>4$. First we treat the case $k_S^2\neq \{0, (d-2)K \}$.
 
\subsubsection{Modes with $\bm{k_S^2\neq \{0, (d-2)K \} }$}

For ${\cal K}^{d-2} = S^{d-2}$ this case corresponds to assuming $l \ge 2$.

Scalar perturbations can be expressed in terms of a single gauge invariant  scalar $\Phi_S$ whose definition can be found in Ref.  \cite{Kodama:2003jz}. This quantity obeys the Kodama-Ishibashi master equation \cite{Kodama:2003jz}.
\begin{equation} 
 \left(  \Box_2 - \frac{V_S}{f} \right) \Phi_S =0 \,, \qquad 
    V_S = \frac{f(r)Q(r)}{16r^2\left[ \gamma + (d-1)(d-2)x/2  \right]^2}  \,,
\label{master:scalar} 
\end{equation}  
where $\Box_2$ is the d'Alembertian defined by $g_{AB}$ and
\begin{eqnarray} 
&&
 \hspace{-0.2cm} x= \frac{r_m^{d-3}}{r^{d-3}} \,,  \qquad  \gamma= k_S^2 - (d-2)K,\nonumber\\      
  &&\hspace{-0.8cm} Q(r) = -\left[
             (d-2)^3d(d-1)^2 x^2 -12(d-2)^2 (d-1)(d-4)\gamma x + 4(d-4)(d-6)\gamma^2 
              \right] \lambda_c r^2 \nonumber\\       
        && \hspace{0.6cm}
            + (d-2)(d-1)\left[4\left((2(d-2)^2-3(d-2)+4\right)\gamma +(d-1)(d-2)(d-4)(d-6)K \right]x^2
\nonumber\\
        &&  \hspace{0.6cm}
           +(d-2)^4(d-1)^2x^3   -12(d-2)\left[(d-6)\gamma+(d-1)(d-2)(d-4) \right] \gamma x +16\gamma^3\nonumber\\
        && \hspace{0.6cm} +4d (d-2)\gamma^2 \,.
\end{eqnarray} 
In terms of the  master variable $\Phi _S$, we find that the algebraically special condition  $\delta \Omega_{ij}=0$ reads
\begin{equation} \label{scalar:dOmegaij}
2r^2 f \,\partial _v\partial _r\Phi _S-\frac{r \left(2 P_Z-Q_X+Q_Y\right)}{4H} \,\partial _v\Phi _S+\frac{r\, f\left(Q_X-Q_Y\right)}{4H}\,\partial _r\Phi _S+\frac{f \left(P_Y-P_X\right)}{16H^2}\,\Phi _S=0 \,.
\end{equation}
where  $P_X,\,P_Y,\,P_Z,\,Q_X,\,Q_Y$, and $H$ are functions of 
$r$ that can be found in Eq.~(3.10)  of \cite{Kodama:2003jz}.

An algebraically special perturbation is a solution of \eqref{master:scalar} and \eqref{scalar:dOmegaij}. 
A combination of these equations yields a necessary but not sufficient condition for an algebraically special perturbation,
\begin{equation} \label{scalar:masterdO}
\partial _r^2\Phi _S-\frac{d-2}{r}\,\frac{(d-4)(d-1) r_m^{d-3}-2r^{d-3} \left[k_S^2-(d-2) K\right]}{(d-2)(d-1) r_m^{d-3}+2r^{d-3} \left[k_S^2-(d-2) K\right]}\,\partial _r\Phi _S+\frac{(d-4)(d-2)}{4r^2} \Phi _S=0\,.
\end{equation}
We will solve this and then substitute into \eqref{master:scalar} and \eqref{scalar:dOmegaij}.
Note the presence of $(d-4)$ factors in \eqref{scalar:masterdO}, namely, in the linear term in $\Phi_S$ and in one of the contributions to $\partial _r\Phi _S$. We thus anticipate that the $d=4$ and $d\geq 5$ cases have distinct properties and we analyze them separately.
\vskip 0.2cm
\noindent{\bf i) Case $\bm{d=4}$: Robinson-Trautman perturbations}
\vskip 0.1cm 

In this case, \eqref{scalar:masterdO} has the general solution 
\be \label{4dRT0}
\Phi _S=\frac{C_1(v)}{\left(2 K-k_S^2\right) \left(3 r_m-r \left(2 K-k_S^2\right)\right)}+C_2(v)
\ee
for arbitrary functions $C_1(v),\,C_2(v)$ (recall that modes with $k_S^2= (d-2)K=2K$ are excluded in the study of this subsection). The requirement that the original equations \eqref{master:scalar} and \eqref{scalar:dOmegaij} are obeyed fixes
\be
C_1(v)=-3 r_m \left(2 K-k_S^2\right) C_2(v), \qquad C_2(v)=A \, e^{-\frac{ k_S^2\left(2 K-k_S^2\right)}{6 r_m}v}
\ee
for some constant $A$. Putting this together gives the solution
\begin{equation} \label{scalar:RobinsonTrautman}
\Phi _S=\frac{A\,r \left(k_S^2-2 K\right)}{r\left(k_S^2-2 K\right)+3 r_m}\,e^{\frac{ k_S^2 \left(k_S^2-2 K\right)}{6 r_m}v}.
\end{equation}
The metric perturbation can be reconstructed in a particular gauge using the linear differential map   $h_{\mu\nu}=h_{\mu\nu}(\Phi_S)$ given in \cite{Kodama:2003jz}. Note that  \eqref{4dRT0}-\eqref{scalar:RobinsonTrautman} are independent of the  cosmological constant, but the corresponding $h_{\mu\nu}(\Phi_S)$ is not.

We have found an infinite class of algebraically special perturbations labelled by the eigenvalue $k_S^2$. These solutions vanish on the past horizon and grow exponentially with $v$.\footnote{Time reversal as discussed at the end of section \ref{sec:AlgSpecial} gives a solution decaying exponentially with $u$ and vanishing on the future horizon. This is the usual form of the Robinson-Trautman solutions.} For $K=1$ and $\Lambda=0$ these perturbations were first identified by Couch and Newman \cite{Couch:1973zc}. They correspond to the linearization around the Schwarzschild solution of the Robinson-Trautman class of algebraically special solutions.

\vskip 0.2cm
\noindent{\bf ii) Case  $\bm{d\geq 5}$:}
\vskip 0.1cm

For $d\geq 5$, the most general solution of \eqref{scalar:masterdO} is
\begin{equation} \label{scalar:master3Sol}
\Phi _S=\frac{2r^{\frac{d-4}{2}} \left[A_1(v)+r A_2(v)\right]}{2 \left[k_S^2-(d-2) K\right]r^{d-3}+(d-2)(d-1) r_m^{d-3}}\,, 
\end{equation}
where $A_1(v),\,A_2(v)$ are arbitrary functions. Plugging \eqref{scalar:master3Sol} into the original equations  \eqref{master:scalar} and \eqref{scalar:dOmegaij} leads to the trivial solution 
\begin{equation} \label{scalar:master3}
A_1(v)= 0=A_2(v) \qquad \Leftrightarrow \qquad  \Phi_S=0.
\end{equation}
Hence there are no algebraically special perturbations of scalar type with $k_S^2\neq \{0, (d-2)K \} $ and $d\ge 5$.

\subsubsection{$\bm{l=1}$ perturbations on $\bm{S^{d-2} }$}

We now take ${\cal K}^{d-2}=S^{d-2}$ and consider the $l=1$ scalar perturbations. For $\lambda_c=0$, $d=4$, it is known that the only such perturbations are pure gauge \cite{Zerilli:1971wd}. Here we will show that the same is true for for any $\lambda_c, d$ by generalizing the $d=4$ argument as presented in Ref. \cite{Martel:2005ir}.

 The equation of motion \eqref{EOMscalarH} and the conditions $\scalar_{ij}=0$ enable us to express all the second derivatives of the scalar harmonic $\scalar $ with $k_S^2= (d-2)$ as a function of $\scalar$ and its first derivatives. Consequently, we can derive the results below without ever introducing the explicit expression for the  $l=1$ scalar harmonics.

An infinitesimal scalar gauge vector $\xi$ can be decomposed in terms of  scalar harmonics as
\begin{equation} \label{scalargauge}
\xi=P_A(x^B) \,\scalar \, dx^A + r \,L(x^B)\, \scalar_{i} \,dx^i\,.
\end{equation}
Choose the gauge $f_{rr}=f_r=H_L=0$ which is preserved by the gauge parameter
\begin{eqnarray} \label{l1gauge}
&& P_v=(f-1) \alpha _1(v)- r \,\alpha _2(v)\,,\nonumber\\      
  && P_r= - \alpha _1(v)\,, \nonumber\\       
        && L=k_S  {\bigl [} \alpha _1(v)+r\,\alpha _2(v) {\bigr ]},
\end{eqnarray} 
for arbitrary functions $\alpha_1(v)$ and $\alpha_2(v)$ of the advanced time $v$. Under this gauge transformation the other components of the metric perturbation transform as  
\begin{eqnarray}  \label{l1fab}
&& f_{vv}\to \: \widetilde{f}_{vv}=f_{vv}+ {\bigl [} \alpha _1(v)+r \,\alpha _2(v)  {\bigr ]} f'-2 (1-f)\alpha _1'(v)-2 r \,\alpha _2'(v)\nonumber\\     
 && f_{vr}\to \: \widetilde{f}_{vr}=f_{vr}- {\bigl [} \alpha _2(v)+\alpha _1'(v) {\bigr ]}\nonumber\\     
 && f_{v\phantom{v}}\to  \: \widetilde{f}_v=f_v+\frac{k_S }{r}{\biggl [}(1-f) \alpha _1(v)+r {\biggl (} \alpha _2(v)+\alpha _1'(v)+r \,\alpha _2'(v) {\biggr )} {\biggr ]},
\end{eqnarray} 
where the $^\prime$ denotes differentiation wrt to the argument of the function.
Consider now the linearized Einstein equation $E_{AB}=0$ (explicitly written in (A.1) of \cite{Kodama:2003jz}) for this perturbation.
The equation $E_{rr}=0$ implies that  $f_{vr}$ is a function of $v$ only and we can set it to zero with the gauge parameter choice $\alpha _2(v)=-\alpha _1'(v)$; see \eqref{l1fab}. In these conditions, equation $E_{ri}=0$ is solved by 
\be
f_v=r\, \gamma_1(v)+\gamma_2(v)/r^{d-2}
\ee 
for arbitrary functions  $\gamma_{1,2}(v)$. Equation $E_{v i}=0$ then implies that 
\be
f_{vv}=\frac{1}{r^{d-4}}\,\gamma _3(v)-\frac{1}{k_S}\left(2\, r \,\gamma _1(v)-\frac{(d-3)}{r^{d-2}}\,\gamma _2(v)+\frac{(d-1)}{r^{d-3}}\,\gamma _2^\prime(v)\right)
\ee
where  $\gamma_3(v)$ is a new arbitrary function of $v$. We must  set  $\gamma_3(v)=0$ to solve  $E_{ii}=0$. Finally, $E_{v v}=0$ implies 
\be
\gamma _1= -\frac{(d-1) r^2 \,\gamma _2^{\prime \prime }+\left[(d-3)(f-1)+r f'\right]\,\gamma _2}{r^{d-1}\left[2(1-f)+r f'\right]}.
\ee
At this point, all components of the Einstein equation are obeyed. Define $\gamma(v)$ by
 \be
 \gamma _2(v)\equiv \frac{r^{d-3}}{d-1}\left[2(1-f)+r f'\,\right]\gamma (v),
 \ee
We conclude that the only $l=1$ scalar modes are 
\begin{eqnarray}  \label{l1fabfinal}
&& f_{vv}=\frac{1}{k_S}{\biggl (}2\, r \,\gamma^{\prime\prime}(v)- {\bigl [} 2(1-f)+r f' \,{\bigr ]}\gamma^\prime(v)+ f'\gamma (v){\biggr )}, \nonumber\\     
 && f_{vr}=0\,,\nonumber\\     
 && f_v=-r \,\gamma^{\prime\prime}(v)+\frac{ 1-f}{r}\, \gamma (v)\,.
\end{eqnarray} 
in the gauge $f_{rr}=f_r=H_L=0$. There is a remaining gauge freedom described by \eqref{l1gauge} and \eqref{l1fab} with $\alpha _2(v)=-\alpha _1'(v)$. Setting  the gauge parameter $\alpha _1(v)=\gamma (v)/k_S$ we find that this gauge transformation yields 
\begin{eqnarray} 
\widetilde{f}_{AB}=0\,,\qquad  \widetilde{f}_{A}=0\,,\qquad \widetilde{H}_L=0\,.
\end{eqnarray} 
Therefore, the $l=1$ scalar modes can indeed be gauged away.

%%%%%%%%%%%%%%%%

%%%%%%%%%%%%%%%%%%
\section{Discussion}\label{Conc}
%%%%%%%%%%%%%%%%%%

To summarize: if there exists a family of vacuum solutions that is type I(a) (or more special) in the classification of \cite{cmpp}, or type $\underline{\bf 22}$ in the 5d classification of \cite{desmet}, and contains the generalized Schwarzschild solution \eqref{gSch}, then the linearization of this family about Schwarzschild yields a solution of the linearized Einstein equation which also satisfies $\delta \Omega_{ij}=0$. We have determined all such perturbations which are regular on (compact) ${\cal K}^{d-2}$. For $d=4$ we find infinite classes of time-dependent perturbations corresponding to those discovered by Couch and Newman. However, for $d>4$, the only perturbations that we find are those corresponding to variation of parameters in the Schwarzschild solution (i.e. the mass, and moduli of ${\cal K}^{d-2}$), or to turning on angular or linear momentum.

One of our motivations was to learn whether there exist new families of algebraically special solutions that contain the Schwarzschild solution. The answer is no, at least if one insists that the family be a {\it smooth} deformation of Schwarzschild. Algebraically special solutions appear to be much scarcer in higher dimensions than in $d=4$, even if one adopts the definition of an algebraically special as one of type I(a), which is weaker than that of most recent work. 

Special cases of our result follow  from previous work on certain classes of solutions admitting {\it multiple} WANDs. Ref. \cite{Godazgar:2009fi} determined all {\it axisymmetric} vacuum spacetimes admitting a multiple WAND. No non-trivial family of solutions containing Schwarzshild was discovered. Recent work of Ref. \cite{Ortaggio:2012cp} implies there exists no non-trivial family of vacuum solutions which admits a multiple WAND which is geodesic\footnote{
This is not restrictive: if there exists a multiple WAND then there exists a geodesic multiple WAND \cite{Durkee:2009nm}.} and {\it non-twisting}, and contains the Schwarzschild solution. To prove this, note that Theorem 1.1 of Ref. \cite{Ortaggio:2012cp} implies that any such family must be Robinson-Trautman.\footnote{The conditions $\det \rho_{ij} \ne 0$ and $\Phi_{ij} \ne 0$ required to reach this conclusion \cite{Ortaggio:2012cp} are satisfied by the Schwarzschild solution hence, by continuity, they will be satisfied for solutions near to Schwarzschild within the family.} For $d=4$, the Robinson-Trautman family contains a large class of time-dependent spacetimes. But the only $d>4$ Robinson-Trautman solution with non-vanishing ``mass function" is the generalized Schwarzschild solution \eqref{gSch} \cite{Podolsky:2006du}.  

It would be desirable to classify algebraically special perturbations without assuming that the perturbation is regular on ${\cal K}^{d-2}$. However, this seems difficult even for $d=4$. Ref. \cite{wald} determined all perturbations of the Kerr solution that satisfy the linearization of the type D condition but without assuming regularity on $S^2$. However, the method relied heavily on the work of Ref. \cite{Kinnersley:1969zza}, in which all type D solution of the vacuum Einstein equation were determined. In higher dimensions we have no analogue of the analysis of Ref. \cite{Kinnersley:1969zza}. 

Our result has significance for the study of {\it general} perturbations of higher-dimensional black holes. Ref. \cite{Durkee:2010qu} showed that $\delta \Omega_{ij}$ is gauge invariant for perturbations of any algebraically special vacuum solution. Since $\delta \Omega_{ij}$ is gauge-invariant and contains the same number of degrees of freedom as a generic metric perturbation, it was suggested that it should be possible to reconstruct a metric perturbation from the corresponding $\delta \Omega_{ij}$ up to the freedom to add modes corresponding to variation of parameters in the background solution. 
Our results show that this is indeed the case for perturbations of Schwarzschild: if two metric perturbations lead to the same $\delta \Omega_{ij}$ then their difference has $\delta \Omega_{ij}=0$ and therefore, by our results, corresponds to a variation of parameters (allowing for a change in angular momenta or a boost).
 
 For $d=4$, $\Lambda=0$ it is known that the Couch-Newman algebraically special perturbations are closely related to quasinormal (QN) modes with purely imaginary frequency (see Ref. \cite{Maassen van den Brink:2000ru} for a discussion). Our results show that this relation does not extend to $\Lambda<0$ (allowing for black holes with $K=0,-1$). Our $\Lambda<0$ generalizations of the Couch-Newman modes have purely imaginary frequencies, but they are not QN modes because they do not satisfy the appropriate (normalizable) boundary conditions at infinity. Conversely, there exist QN modes with purely imaginary frequency \cite{Cardoso:2001bb} but these frequencies differ from those of our algebraically special modes. Hence for $\Lambda<0$ it appears that there is no relation between algebraically special perturbations and QN modes with purely imaginary frequency. For $d>4$, $\Lambda=0$ we have found no time-dependent algebraically special perturbations and numerical work indicates that purely imaginary QN modes also do not exist \cite{Cardoso:2003vt}. However, we know of no reason why these observations should be related.

 %%%%%%%%%%%%%%%%
 \subsubsection*{Acknowledgments}
We are grateful to Mahdi Godazgar for useful discussions, and for having reproduced independently equation \eqref{dOmegaij} and to Vojtech Pravda and Alena Pravdova for the argument in footnote \ref{pravda}. 
 We also thank Jorge Santos for useful discussions. OD thanks the Yukawa Institute for Theoretical Physics (YITP) at Kyoto University, where part of this work was completed during the YITP-T-11-08 programme  ``Recent advances in numerical and analytical methods for black hole dynamics", and the  participants of the workshops ``The Holographic Way: String Theory, Gauge Theory and Black Holes", Nordita (Sweden), ``Spanish Relativity Meeting in Portugal",  ``Exploring AdS-CFT Dualities in Dynamical Settings", Perimeter Institute (Canada), and ``Numerical Relativity and High Energy Physics", Madeira (Portugal) for discussions.  HSR is supported by a Royal Society University Research Fellowship and by European Research Council grant no. ERC-2011-StG 279363-HiDGR. 
%%%%%%%%%%%%%%%%%%%%%%%%%%%%%%%%%%%%%%%%%%%%%%%%%
%%%%%%%%%%%%%%%%%%%%%%%%%%%%%%%%%%%%%%%%%%%%%%%%%

 \appendix
\renewcommand{\theequation}{A.\arabic{equation}}

%%%%%%%%%%%%%%%%%%%%%%%%%%%%%%%%%%%%%%%%%%%%%
\section{Perturbations of a type D Einstein spacetime}\label{sec:algSpecialPert}
%%%%%%%%%%%%%%%%%%%%%%%%%%%%%%%%%%%%%%%%%%%%%%%%

We use the higher-dimensional generalization of the GHP formalism \cite{ghp}, which was developed in Ref. \cite{Durkee:2010xq}. We ask the reader to see section 2 of \cite{Durkee:2010xq} for the GHP notation and properties required to follow the derivation of this Appendix.

We are interested in linearized gravitational perturbations of a type D Einstein spacetime. Such a geometry is defined  by the conditions
\begin{eqnarray}
&&  \Omega^{(0)}_{ij}=0,\quad\Omega^{\prime\, (0)}_{ij}=0,\quad\Psi^{(0)}_{ijk}=0,\quad\Psi^{\prime\, (0)}_{ijk}=0,\quad\Psi^{(0)}_{i}=0,\quad\Psi^{\prime\, (0)}_{i}=0,\label{WeyltypeD}\\
&& R_{\alpha\beta}=\frac{2 \lambda_c}{d-2}\, g_{\alpha\beta}. \label{Einsteinspace}
\end{eqnarray}

For a quantity $X$, we shall write $X=X^{(0)}+\delta X$ where $X^{(0)}$ is the value in the background spacetime and $\delta X$ is the perturbation.  
As described in Section \ref{sec:AlgSpecial}, we want to find the expression for $\delta\Omega_{ij}$ (the perturbation in $\Omega_{ij}$) which is gauge invariant under infinitesimal coordinate and basis transformations. An algebraically special perturbation obeys  \eqref{algSpecPert}, i.e. $\delta\Omega_{ij}=0$.

The variation $\delta\Omega_{ij}$ includes two main contributions, one that comes from the variation of the basis under a perturbation, and the other that is due to the variation of the Weyl tensor itself, 
\begin{eqnarray}
\delta\Omega_{ij}\equiv \delta C_{0i0j}=\delta\left( \lb^\alpha \mb{i}^\beta \lb^\mu \mb{j}^\nu C_{\alpha \beta \mu \nu } \right)=2 C_{a(i|0|j)} \delta\lb^a+\left(\delta C\right)_{0i0j}
\label{initialdOij}
\end{eqnarray}
where in the last equality we used the symmetries of the Weyl tensor and we assumed that the background is Petrov type D. We use the notation $\left(\delta C\right)_{0i0j}\equiv \lb^\alpha \mb{i}^\beta \lb^\mu \mb{j}^\nu \delta \left( C_{\alpha \beta \mu \nu}\right) $. Also, we use $\delta \lb^a \equiv \delta \left( \lb^a \right)$ to represent the variation of the vector $ \lb^a$. Then, $\delta \lb_a\equiv \eta_{ab}\delta\lb^b$ and recall that $\delta\lb^a=\eb^a_{\phantom{x}\mu} \delta\lb^\mu$. \footnote{The variation of the covector $\lb_{a}$ is then related to the variation of the dual vector through $\delta \left(  \lb_a\right)= \delta \left( \eta_{ab}  \lb^b \right)=  \delta \eta_{ab}  \lb^b+  \eta_{ab}  \delta \lb^b \equiv  h_{ab}  \lb^b+   \delta \lb_a$.}
 To find the variation of a vector of the null basis we vary the expression for the background metric in terms of the null basis vectors,
\begin{equation}
-h^{\mu\nu}\equiv\delta g^{\mu\nu}=\delta\left( 2\lb^{(\mu}\nb^{\nu )}+\delta_{ij}\mb{i}^{\mu}\mb{j}^{\nu}\right),
\label{g:basis}
\end{equation}
 which, for example, allows to find that the null basis components of the variation of $\lb$ are
 \begin{eqnarray}
&& \delta \ell^a=\eb^a_{\phantom{x}\mu} \delta\lb^\mu=-h_0^{\phantom{x}a} - \nb^a \delta\lb_0 - \lb^a \delta\nb_0 -\delta_{ij}\mb{i}^a \delta\mb{j}_0 \nonumber\\
   && \hspace{1cm} \rightarrow\quad 2\delta\lb^0=-h_{01}-\delta\nb_0\,, \quad 2\delta\lb^1=-h_{00}\,,\quad \delta \lb^i=-h_{0i}-\delta\mb{i}_0\,,
\label{deltabasis}
\end{eqnarray}
where we used the orthogonality conditions \eqref{nullbasis} for the null basis.

At this stage, using \eqref{deltabasis} together with the symmetries of the Weyl tensor,  Table 2 of  \cite{Durkee:2010xq} and the relations \eqref{WeyltypeD} valid for a type D background, we can rewrite \eqref{initialdOij} as 
\begin{equation}
\delta\Omega_{ij}=2 C_{1(i|0|j)} \delta\lb^1+\left(\delta C\right)_{0i0j}=-h_{00}\Phi^{\rm S}_{ij}+\left(\delta C\right)_{0i0j}
\label{initialdOij:2}
\end{equation}
To compute  $\left(\delta C\right)_{0i0j}$, we vary the expression that decomposes the Weyl tensor in terms of the Riemann and Ricci tensors and Ricci scalar, namely,
\begin{equation}
C_{\alpha \beta \mu \nu }=R_{\alpha \beta \mu \nu }-\frac{1}{d-2}\left(g_{\alpha \mu }R_{\beta \nu }-g_{\beta \mu }R_{\alpha \nu }-g_{\alpha \nu }R_{\beta \mu }+g_{\beta \nu }R_{\alpha \mu }\right)+\frac{\left(g_{\alpha \mu }g_{\beta \nu }-g_{\alpha \nu }g_{\beta \mu }\right)R}{(d-1)(d-2)}.
\label{weyltensor}
\end{equation} 
This variation is accomplished if we use the expression for the variation of the affine connection, $\delta\Gamma^\mu_{\phantom{x}\alpha\beta}$, Palatini's identity for the variation of the Riemann tensor $\delta R^\alpha_{\phantom{x}\beta\mu\nu}$, and the definitions of the (variation of the) Ricci tensor $\delta R_{\alpha\beta}$ and Ricci scalar  $\delta R$,
\begin{eqnarray}
&& \hspace{-1.8cm}\delta\Gamma^\alpha_{\phantom{x}\mu\nu}=\frac{1}{2}\left[ g^{\alpha\beta}\left( \partial_\mu h_{\nu\beta}
+ \partial_\nu h_{\mu\beta} -\partial_\beta h_{\mu\nu}\right) -h^{\alpha\beta}\left( \partial_\mu g_{\nu\beta}
+ \partial_\nu g_{\mu\beta} -\partial_\beta g_{\mu\nu}\right)\right],\nonumber\\
&&   \hspace{-1.8cm} \delta R^\alpha_{\phantom{x}\beta\mu\nu}=\nabla_\mu\left( \delta\Gamma^\alpha_{\phantom{x}\beta\nu} \right)-\nabla_\nu\left( \delta\Gamma^\alpha_{\phantom{x}\beta\mu} \right)\,,\qquad  \delta R_{\alpha \beta}= \delta R^\mu_{\phantom{x}\alpha\mu\beta}\,,\qquad  \delta R= \delta\left( g^{\alpha \beta}R_{\alpha \beta}\right).
\label{deltaRie}
\end{eqnarray}
In these expressions, $g$ is the background metric,  $\nabla$ is the associated background Levi-Civita connection, and we will henceforth take the background to be an Einstein spacetime \eqref{Einsteinspace}.

In these conditions, at this point we can write
\begin{eqnarray}
&&\hspace{-1cm} 2\left(\delta C\right)_{0i0j}=\lb^a\mb{(i}^b\mb{j)}^d D\left( 2\nabla_b h_{ad} - \nabla_a h_{bd} \right) -\lb^a\mb{(i}^b\lb^c \delta_{j)} \nabla_b h_{ac}+2h_{00}\Phi^{\rm S}_{ij}\nonumber\\
&& \hspace{1.2cm}+\frac{2\Lambda}{d-2}h_{00}\delta_{ij}-\frac{1}{d-2}\delta_{ij}\lb^a\lb^c\left( 2\nabla_f\nabla_c h^f_{\phantom{a}a}-\nabla^2h_{ac}-\nabla_c\nabla_a h \right),
\label{dC0i0j}
\end{eqnarray}
where we use the notation $ D \equiv \lb \cdot \nabla, \quad \Delta \equiv \nb \cdot \nabla \eqand \del_i \equiv \mb{i} \cdot \nabla$ for the components of the covariant derivative operator in the null frame, and recall that $\nabla_a=\eb_a^{\phantom{\mu}\mu}\nabla_\mu$.
To proceed we  make use of several definitions/notation/properties of the GHP formalism listed in section 2 of \cite{Durkee:2010xq}. More concretely,  we need to use the projection of a tensor into the null basis $ T_{ab...c} = e^{\phantom{a}\mu}_{a} e^{\phantom{a}\nu}_{b} ... e^{\phantom{a}\alpha}_{c}\, T_{\mu\nu...\alpha}$; the components of the covariant derivative in the null frame; the covariant derivative of the basis vectors,
 \begin{equation}\label{LNM}
 L_{ab} = \nabla_b l_a,\qquad
  N_{ab} = \nabla_b n_a,\qquad
  \M{i}_{ab} = \nabla_b m_{(i)a},
\end{equation}
and associated notation listed in Table \ref{tab:weights}; 
and the identities (which follow from the orthogonality properties of the basis vectors)
\begin{eqnarray}\label{eqn:ident}
 && \hspace{-1.5cm} N_{0a} + L_{1a} = 0, \quad \M{i}_{0a} + L_{ia} = 0,
  \quad \M{i}_{1a} + N_{ia} = 0, \quad \M{i}_{ja} + \M{j}_{ia} = 0, \nonumber\\
&& \hspace{-1.5cm}  L_{0a} = N_{1a} = \M{i}_{ia} = 0\,, \qquad  L_{10} = -N_{00}, \quad L_{11} = -N_{01} \eqand L_{1i} = -N_{0i}\,.
\end{eqnarray}
We also need to use the GHP derivative operators $\tho$, $\tho'$, $\eth_i$ that map GHP scalars to GHP scalars.  They act on a GHP scalar $T_{i_1 i_2...i_s}$ of spin $s$ and boost weight $b$ as:
\begin{eqnarray}  \label{GHPder} 
  \tho T_{i_1 i_2...i_s} 
      &\equiv & (\lb\! \cdot\! \partial ) T_{i_1 i_2...i_s} - b L_{10} T_{i_1 i_2...i_s}
                + \sum_{r=1}^s \M{k}_{i_r 0} T_{i_1...i_{r-1} k i_{r+1}...i_s},\\
  \tho' T_{i_1 i_2...i_s} 
      &\equiv & (\nb\! \cdot\! \partial ) T_{i_1 i_2...i_s} - b L_{11} T_{i_1 i_2...i_s}
                + \sum_{r=1}^s \M{k}_{i_r 1} T_{i_1...i_{r-1} k i_{r+1}...i_s},\\
  \eth_i T_{j_1 j_2...j_s} 
      &\equiv & (\mb{i}\! \cdot\! \partial )T_{j_1 j_2...j_s} - b L_{1i} T_{j_1 j_2...j_s}
                + \sum_{r=1}^s \M{k}_{j_r i} T_{j_1...j_{r-1} k j_{r+1}...j_s}.
\end{eqnarray}
These GHP derivative operators have properties that we use through our computation, namely they  are GHP covariant, obey the Leibniz rule and are metric for $\delta_{ij}$ and satisfy the commutator relations listed in \cite{Durkee:2010xq}. Our guideline and final target is to write $\delta\Omega_{ij}$ uniquely as a function of GHP scalar quantities, listed in Table \ref{tab:weights} and \ref{tab:weightsH}, and their GHP derivatives.  In the end of this process we find that an algebraically special perturbation must obey  
 
\begin{eqnarray}
&&\hspace{-0.8cm} \delta\Omega_{ij}={\biggl [} -\frac{1}{2}\eth_{(i}\eth_{j)}  +\tau^\prime_{(i}\eth_{j)}+\frac{1}{2}\rho^\prime_{(ij)}\tho
  -\frac{1}{2}\rho_{(ij)}\tho^\prime +\kappa_{(i}\kappa^\prime_{j)} +\rho_{k(i}\rho^\prime_{|k|j)}+ \left(  
  \tho\rho^\prime_{(ij)}\right)\nonumber\\
&&\hspace{0.3cm} -\frac{\delta_{ij}}{d-2}{\biggl (}    
  -\frac{1}{2}\eth_{k}\eth_{k}  +\tau^\prime_{k}\eth_{k}+\frac{1}{2}\rho^\prime \tho
  -\frac{1}{2}\rho\tho^\prime +\kappa_{k}\kappa^\prime_{k} +\rho_{lk}\rho^\prime_{lk}+ \left( \tho\rho^\prime\right) {\biggr )} {\biggr ]}h_{00} \nonumber\\
&&\hspace{-0.2cm}+{\biggl [} \rho_{(ij)}\tho +\kappa_{(i}\eth_{j)}+\kappa_{(i}\tau_{j)}-\kappa_{(i}\tau^\prime_{j)} 
    +\rho_{k(i}\rho_{|k|j)} + \left( \tho\rho_{(ij)}\right)\nonumber\\
&&\hspace{0.3cm}
  -\frac{\delta_{ij}}{d-2}{\biggl (}   \rho\tho +\kappa_{k}\eth_{k}-2\kappa_{k}\tau^\prime_{k} -2\rho_{l k}\rho_{[kl]} + \left(  \eth_k\kappa_{k}\right)   {\biggr )}{\biggr ]}h_{01}\nonumber\\
&&\hspace{-0.2cm}
+{\biggl [} 2\rho_{k(i}\eth_{j)} -\kappa_{(i}\rho^\prime_{|k|j)}-2\tau^\prime_{(i}\rho_{|k|j)}+\tau_k\rho_{(ij)}
 + \left(  \eth_{(i}\rho_{|k|j)}\right)   -\frac{\delta_{ij}}{d-2}{\biggl (} \eth_k\tho-2\tau^\prime_k\tho+\left(2\rho_{kl}-\rho_{lk}\right)\eth_l  \nonumber\\
&&\hspace{0.3cm}  
 +\rho\left( \tau_k-\tau^\prime_k \right) +\kappa_l\rho^\prime_{kl}-\tau^\prime_l\rho_{lk}-\rho^\prime\kappa_k-\rho_{kl}\left(2\tau^\prime_l-\tau_l \right)
+{\bigl(} \tho (\tau_k-\tau^\prime_k){\bigr )} +\left( \eth_l\rho_{kl}\right) -\left( \tho^\prime k_k\right) {\biggr )}
{\biggr ]}h_{0k}\nonumber\\
&&\hspace{-0.2cm}-{\biggl [} \kappa_k\rho_{(ij)}+\kappa_{(i}\rho_{|k|j)} -\frac{\delta_{ij}}{d-2}{\biggl (}    
 2\kappa_k\tho+\rho\kappa_k+\kappa_l\rho_{lk}+\left( \tho \kappa_k\right) {\biggr )}  {\biggr ]}h_{1k}-{\biggl [}\kappa_{(i}\kappa_{j)}-\frac{\delta_{ij} \kappa_k\kappa_k }{d-2}{\biggr ]}h_{11}\nonumber\\
&&\hspace{-0.2cm}-{\biggl [}  \rho_{k(i}\rho_{|l|j)}  +\frac{\delta_{ij}}{d-2}{\biggl (}    
-\rho_{kl}\tho-\kappa_l\eth_k+2\kappa_k\tau^\prime_l+2\rho_{lm}\rho_{[mk]}-\left( \eth_k \kappa_l\right) {\biggr )}  {\biggr ]}h_{kl}\nonumber\\
&&\hspace{-0.2cm}+{\biggl [} \tho\eth_{(i} -\tau^\prime_{(i}\tho+ \kappa_{(i}\tho^\prime+\kappa_k\left(  \rho^\prime_{(i|k|} \!-\!  \rho^\prime_{k(i}\right)-\left( \tho\tau^\prime_{(i}\right)\!\!{\biggr ]}h_{j)0}
 -{\biggl [}  \kappa_{(i}\tho-\kappa_k\left(  \rho_{(i|k|} \!-\! \rho_{k(i}\right)+\left( \tho\kappa_{(i}\right)\!\! {\biggr ]}h_{j)1}\nonumber\\
&&\hspace{-0.2cm}-{\biggl [}\kappa_k\eth_{(i}+\rho_{k(i}\tho-\left(\tau^\prime_k-\tau_k\right)\kappa_{(i}+\left( \tho\rho_{k(i}\right) {\biggr ]}h_{j)k}-\frac{1}{2}{\biggl [}\tho\tho-\kappa_l\eth_l{\biggr ]} {\biggl (} h_{(ij)}-\frac{\delta_{ij}}{d-2}h_{kk} {\biggr )}=0.\nonumber\\
&&
\label{dOmegaij}
\end{eqnarray}
As required, $\delta\Omega_{ij}$ has the symmetries of $\Omega_{ij}$ $-$ it is symmetric and traceless $-$ and it is a spin $2$, boost weight $2$ GHP scalar. Recall, the conditions for the validity of \eqref{dOmegaij}: the background must be type D and an Einstein spacetime and no gauge choice was made in the derivation of this expression.

\begin{table}[ht]
 \begin{center}
   \begin{tabular}{|c|c|c|c|l|}
    \hline Quantity & Notation & Boost weight $b$ & Spin $s$ & Interpretation\\ [1mm]\hline
    $L_{ij}$  & $\rho_{ij}$  & 1  & 2 & expansion, shear and twist of $\lb$\\[1mm]
     $L_{ii}$  & $\rho=\rho_{ii}$  & 1  & 0 & expansion of $\lb$\\[1mm]
    $L_{i0}$  & $\kap_{i}$   & 2  & 1 & non-geodesity of $\lb$\\[1mm]
    $L_{i1}$  & $\tau_{i}$   & 0  & 1 & transport of $\lb$ along $n$\\[1mm]
    $N_{ij}$  & $\rho'_{ij}$ & -1 & 2 & expansion, shear and twist of $n$\\[1mm]
     $N_{ii}$  & $\rho'=\rho'_{ii}$ & -1 & 0 & expansion of $n$\\[1mm]
     $N_{i1}$  & $\kap'_{i}$  & -2 & 1 & non-geodesity of $n$\\[1mm]
    $N_{i0}$  & $\tau'_{i}$  & 0  & 1 & transport of $n$ along $l$\\[1mm]\hline
  \end{tabular}
  \caption{\label{tab:weights}GHP scalars constructed from first derivatives of the null basis vectors.}
 \end{center}
\end{table}

\begin{table}[ht]
 \begin{center}
   \begin{tabular}{|c|c|c||||c|c|c|}
    \hline Quantity & Boost weight  $b$  & Spin $s$  & Quantity & Boost weight $b$ & Spin $s$ \\ [1mm]\hline
    $h_{00}$  & 2  & 0  & $h_{0j}$  &  0 & 0 \\[1mm]
    $h_{01}$  & 0  & 0  & $ h_{1j}$ & -1 & 1 \\[1mm]
    $h_{11}$  &-2  & 0  & $h_{ij}$   &  0 & 2 \\[1mm]\hline
  \end{tabular}
  \caption{\label{tab:weightsH} Boost weight $b$ and spin $s$ of the GHP scalars built out of the  metric  perturbation.}
 \end{center}
\end{table}

%%%%%%%%%%%%%%%%%%%%%%%%%%%%%%%%%%%%%%%%%%%%%
%%%%%%%%%%%%%%%%%%%%%%%%%%%%%%%%%%%%%%%%%%%%%
%%%%%%%%%%%%%%%%%%%%%%%%%%%%%%%%%%%%%%%%%%%%%
%%%%%%%%%%%%%%%%%%%%%%%%%%%%%%%%%%%%%%%%%%%%%


\begin{thebibliography}{10}

%\cite{Kerr:1963ud}
\bibitem{Kerr:1963ud} 
  R.~P.~Kerr,
  ``Gravitational field of a spinning mass as an example of algebraically special metrics,''
  Phys.\ Rev.\ Lett.\  {\bf 11}, 237 (1963).
  %%CITATION = PRLTA,11,237;%%

%\cite{Kinnersley:1969zza}
\bibitem{Kinnersley:1969zza} 
  W.~Kinnersley,
  ``Type D Vacuum Metrics,''
  J.\ Math.\ Phys.\  {\bf 10}, 1195 (1969).
  %%CITATION = JMAPA,10,1195;%%

\bibitem{desmet}
  P.~-J.~De Smet,
  ``Black holes on cylinders are not algebraically special,''
  Class.\ Quant.\ Grav.\  {\bf 19}, 4877 (2002)
  [hep-th/0206106].
  %%CITATION = HEP-TH/0206106;%%

\bibitem{cmpp}
A.~Coley, R.~Milson, V.~Pravda, and A.~Pravdov\'a, ``Classification of the Weyl Tensor in Higher Dimensions",
\newblock Class. Quant. Grav. {\bf 21}, L35 (2004), arXiv:gr-qc/0401008.

\bibitem{taghavi} 
  A.~Taghavi-Chabert,
  ``Optical structures, algebraically special spacetimes, and the Goldberg-Sachs theorem in five dimensions,''
  Class.\ Quant.\ Grav.\  {\bf 28}, 145010 (2011)
  [arXiv:1011.6168 [gr-qc]].
  %%CITATION = ARXIV:1011.6168;%%

%\cite{Durkee:2010xq}
\bibitem{Durkee:2010xq}
  M.~Durkee, V.~Pravda, A.~Pravdova and H.~S.~Reall,
  ``Generalization of the Geroch-Held-Penrose formalism to higher dimensions,''
  Class.\ Quant.\ Grav.\  {\bf 27} (2010) 215010
  [arXiv:1002.4826 [gr-qc]].
  %%CITATION = ARXIV:1002.4826;%%

\bibitem{reall} 
  H.~S.~Reall, "Algebraically special solutions in higher dimensions", in ``Black holes in higher dimensions", ed. G. Horowitz, Cambridge University Press (2012);
  arXiv:1105.4057 [gr-qc].
  %%CITATION = ARXIV:1105.4057;%%

%\cite{Ortaggio:2012jd}
\bibitem{Ortaggio:2012jd} 
  M.~Ortaggio, V.~Pravda and A.~Pravdova,
  %``Algebraic classification of higher dimensional spacetimes based on null alignment,''
  Class.\ Quant.\ Grav.\  {\bf 30}, 013001 (2013)
  [arXiv:1211.7289 [gr-qc]].
  %%CITATION = ARXIV:1211.7289;%%

\bibitem{myersperry} R. C. Myers and M. J. Perry, ``Black holes in higher dimensional space-times,"
Annals Phys. {\bf 172}, 304 (1986).

%\cite{Durkee:2010qu}
\bibitem{Durkee:2010qu}
  M.~Durkee and H.~S.~Reall,
  ``Perturbations of higher-dimensional spacetimes,''
  Class.\ Quant.\ Grav.\  {\bf 28} (2011) 035011
  [arXiv:1009.0015 [gr-qc]].
  %%CITATION = ARXIV:1009.0015;%%

%\cite{Teukolsky:1972my}
\bibitem{Teukolsky:1972my} 
  S.~A.~Teukolsky,
  ``Rotating black holes - separable wave equations for gravitational and electromagnetic perturbations,''
  Phys.\ Rev.\ Lett.\  {\bf 29}, 1114 (1972).
  %%CITATION = PRLTA,29,1114;%%

\bibitem{wald}
R.M. Wald,
"On perturbations of a Kerr black hole,"
J. Math. Phys. {\bf 14}, 1453 (1973). 


%\cite{Couch:1973zc}
\bibitem{Couch:1973zc}
  W.~E.~Couch and E.~T.~Newman,
  ``Algebraically special perturbations of the Schwarzschild metric,''
  J.\ Math.\ Phys.\  {\bf 14} (1973) 285.
  %%CITATION = JMAPA,14,285;%%

%\cite{Robinson:1962zz}
\bibitem{Robinson:1962zz} 
  I.~Robinson and A.~Trautman,
  ``Some spherical gravitational waves in general relativity,''
  Proc.\ Roy.\ Soc.\ Lond.\ A {\bf 265}, 463 (1962).
  %%CITATION = PRSLA,A265,463;%%

\bibitem{exact}
H.~Stephani, D.~Kramer, M.~MacCallum, C.~Hoenselaers, and E.~Herlt,
\newblock {\em {Exact solutions of Einstein's field equations}} (Cambridge University
  Press, 2003).
  
%\cite{Kodama:2003jz}
\bibitem{Kodama:2003jz}
  H.~Kodama, A.~Ishibashi, ``A Master equation for gravitational perturbations of maximally symmetric black holes in higher dimensions,'' Prog.\ Theor.\ Phys.\  {\bf 110 } (2003)  701-722.
  [hep-th/0305147].
  
  %\cite{Gibbons:2002pq}
\bibitem{Gibbons:2002pq}
  G.~Gibbons and S.~A.~Hartnoll,
  ``A gravitational instability in higher dimensions,''
  Phys.\ Rev.\ D {\bf 66} (2002) 064024
  [hep-th/0206202].
  %%CITATION = HEP-TH/0206202;%%

%\cite{Kodama:2003kk}
\bibitem{Kodama:2003kk}
  H.~Kodama and A.~Ishibashi,
  ``Master equations for perturbations of generalized static black holes with charge in higher dimensions,''
  Prog.\ Theor.\ Phys.\  {\bf 111} (2004) 29
  [hep-th/0308128].
  %%CITATION = HEP-TH/0308128;%%

%\cite{Kodama:2000fa}
\bibitem{Kodama:2000fa}
  H.~Kodama, A.~Ishibashi and O.~Seto,
  ``Brane world cosmology: Gauge invariant formalism for perturbation,''
  Phys.\ Rev.\ D {\bf 62} (2000) 064022
  [hep-th/0004160].
  %%CITATION = HEP-TH/0004160;%%

%\cite{Kodama:2004kz}
\bibitem{Kodama:2004kz} 
  H.~Kodama,
  ``Perturbative uniqueness of black holes near the static limit in arbitrary dimensions,''
  Prog.\ Theor.\ Phys.\  {\bf 112}, 249 (2004)
  [hep-th/0403239].
  %%CITATION = HEP-TH/0403239;%%

%\cite{Kodama:2008wf}
\bibitem{Kodama:2008wf} 
  H.~Kodama,
  ``Accelerating a Black Hole in Higher Dimensions,''
  Prog.\ Theor.\ Phys.\  {\bf 120}, 371 (2008)
  [arXiv:0804.3839 [hep-th]].
  %%CITATION = ARXIV:0804.3839;%%

%\cite{Frolov:2003en}
\bibitem{Frolov:2003en} 
  V.~P.~Frolov and D.~Stojkovic,
  ``Particle and light motion in a space-time of a five-dimensional rotating black hole,''
  Phys.\ Rev.\ D {\bf 68}, 064011 (2003)
  [gr-qc/0301016].
  %%CITATION = GR-QC/0301016;%%

%\cite{Hamamoto:2006zf}
\bibitem{Hamamoto:2006zf} 
  N.~Hamamoto, T.~Houri, T.~Oota and Y.~Yasui,
  ``Kerr-NUT-de Sitter curvature in all dimensions,''
  J.\ Phys.\ A {\bf 40}, F177 (2007)
  [hep-th/0611285].
  %%CITATION = HEP-TH/0611285;%%

%\cite{DeSmet:2003kt}
\bibitem{DeSmet:2003kt} 
  P.~-J.~De Smet,
  ``The Petrov type of the five-dimensional Myers-Perry metric,''
  Gen.\ Rel.\ Grav.\  {\bf 36}, 1501 (2004)
  [gr-qc/0312021].
  %%CITATION = GR-QC/0312021;%%

\bibitem{godazgar} 
  M.~Godazgar,
  ``Spinor classification of the Weyl tensor in five dimensions,''
  Class.\ Quant.\ Grav.\  {\bf 27}, 245013 (2010)
  [arXiv:1008.2955 [gr-qc]].
  %%CITATION = ARXIV:1008.2955;%%

%\cite{Bais:1984xb}
\bibitem{Bais:1984xb} 
  F.~A.~Bais and P.~Batenburg,
  ``A New Class Of Higher Dimensional Kaluza-klein Monopole And Instanton Solutions,''
  Nucl.\ Phys.\ B {\bf 253}, 162 (1985).
  %%CITATION = NUPHA,B253,162;%%

%\cite{Mann:2003zh}
\bibitem{Mann:2003zh} 
  R.~B.~Mann and C.~Stelea,
  ``Nuttier (A)dS black holes in higher dimensions,''
  Class.\ Quant.\ Grav.\  {\bf 21}, 2937 (2004)
  [hep-th/0312285].
  %%CITATION = HEP-TH/0312285;%%

%\cite{Lu:2004ya}
\bibitem{Lu:2004ya} 
  H.~Lu, D.~N.~Page and C.~N.~Pope,
  ``New inhomogeneous Einstein metrics on sphere bundles over Einstein-Kahler manifolds,''
  Phys.\ Lett.\ B {\bf 593}, 218 (2004)
  [hep-th/0403079].
  %%CITATION = HEP-TH/0403079;%%

%\cite{Ortaggio:2012cp}
\bibitem{Ortaggio:2012cp} 
  M.~Ortaggio, V.~Pravda and A.~Pravdova,
  ``On the Goldberg-Sachs theorem in higher dimensions in the non-twisting case,''
  arXiv:1211.2660 [gr-qc].
  %%CITATION = ARXIV:1211.2660;%%


%\cite{Carter:1968ks}
\bibitem{Carter:1968ks}
  B.~Carter,
  ``Hamilton-Jacobi and Schrodinger separable solutions of Einstein's
  equations,''
  Commun.\ Math.\ Phys.\  {\bf 10} (1968) 280.
  %%CITATION = CMPHA,10,280;%%

%\cite{Hawking:1998kw}
\bibitem{Hawking:1998kw}
  S.~W.~Hawking, C.~J.~Hunter and M.~Taylor,
  ``Rotation and the AdS/CFT correspondence,''
  Phys.\ Rev.\  D {\bf 59}, 064005 (1999)
  [arXiv:hep-th/9811056].
  %%CITATION = PHRVA,D59,064005;%%

%\cite{Gibbons:2004uw}
\bibitem{Gibbons:2004uw}
  G.~W.~Gibbons, H.~Lu, D.~N.~Page and C.~N.~Pope,
  ``The general Kerr-de Sitter metrics in all dimensions,''
  J.\ Geom.\ Phys.\  {\bf 53} (2005) 49
  [arXiv:hep-th/0404008].
  %%CITATION = JGPHE,53,49;%%


\bibitem{yano}
K. Yano and T. Nagano, "Einstein spaces admitting a one-parameter group of conformal transformations", Annals of Mathematics {\bf 69}, 451 (1959). 

%\cite{Zerilli:1971wd}
\bibitem{Zerilli:1971wd}
  F.~J.~Zerilli,
  ``Gravitational field of a particle falling in a schwarzschild geometry analyzed in tensor harmonics,''
  Phys.\ Rev.\ D {\bf 2} (1970) 2141.
  %%CITATION = PHRVA,D2,2141;%%

%\cite{Martel:2005ir}
\bibitem{Martel:2005ir}
  K.~Martel and E.~Poisson,
  ``Gravitational perturbations of the Schwarzschild spacetime: A Practical covariant and gauge-invariant formalism,''
  Phys.\ Rev.\ D {\bf 71} (2005) 104003
  [gr-qc/0502028].
  %%CITATION = GR-QC/0502028;%%

%\cite{Godazgar:2009fi}
\bibitem{Godazgar:2009fi} 
  M.~Godazgar and H.~S.~Reall,
  ``Algebraically special axisymmetric solutions of the higher-dimensional vacuum Einstein equation,''
  Class.\ Quant.\ Grav.\  {\bf 26}, 165009 (2009)
  [arXiv:0904.4368 [gr-qc]].
  %%CITATION = ARXIV:0904.4368;%%


%\cite{Durkee:2009nm}
\bibitem{Durkee:2009nm} 
  M.~Durkee and H.~S.~Reall,
  ``A Higher-dimensional generalization of the geodesic part of the Goldberg-Sachs theorem,''
  Class.\ Quant.\ Grav.\  {\bf 26}, 245005 (2009)
  [arXiv:0908.2771 [gr-qc]].
  %%CITATION = ARXIV:0908.2771;%%


%\cite{Podolsky:2006du}
\bibitem{Podolsky:2006du} 
  J.~Podolsky and M.~Ortaggio,
  ``Robinson-Trautman spacetimes in higher dimensions,''
  Class.\ Quant.\ Grav.\  {\bf 23}, 5785 (2006)
  [gr-qc/0605136].
  %%CITATION = GR-QC/0605136;%%

%\cite{Maassen van den Brink:2000ru}
\bibitem{Maassen van den Brink:2000ru} 
  A.~Maassen van den Brink,
  %``Analytic treatment of black hole gravitational waves at the algebraically special frequency,''
  Phys.\ Rev.\ D {\bf 62}, 064009 (2000)
  [gr-qc/0001032].
  %%CITATION = GR-QC/0001032;%%

%\cite{Cardoso:2001bb}
\bibitem{Cardoso:2001bb} 
  V.~Cardoso and J.~P.~S.~Lemos,
  ``Quasinormal modes of Schwarzschild anti-de Sitter black holes: Electromagnetic and gravitational perturbations,''
  Phys.\ Rev.\ D {\bf 64}, 084017 (2001)
  [gr-qc/0105103].
  %%CITATION = GR-QC/0105103;%%

%\cite{Cardoso:2003vt}
\bibitem{Cardoso:2003vt} 
  V.~Cardoso, J.~P.~S.~Lemos and S.~Yoshida,
  ``Quasinormal modes of Schwarzschild black holes in four-dimensions and higher dimensions,''
  Phys.\ Rev.\ D {\bf 69}, 044004 (2004)
  [gr-qc/0309112].
  %%CITATION = GR-QC/0309112;%%

\bibitem{ghp}
R.~{Geroch}, A.~{Held}, and R.~{Penrose}, ``A spacetime calculus based on pairs of null directions",
\newblock Journal of Mathematical Physics {\bf 14}, 874 (1973).


%%%%%%%%%%%%%%%%%%%%%%%%%
\end{thebibliography}
\end{document}